\documentclass[traditabstract]{aa}

\usepackage{txfonts}
\usepackage{graphicx}

\begin{document}

\title{A mature cluster with X-ray emission at $z=2.07$}

\author{R. Gobat\inst{1}
\and E. Daddi\inst{1} 
\and M. Onodera\inst{2} 
\and A. Finoguenov\inst{3} 
\and A. Renzini\inst{4} 
\and N. Arimoto\inst{5,6} 
\and R. Bouwens\inst{7} 
\and M. Brusa\inst{3} 
\and R.-R. Chary\inst{8} 
\and A. Cimatti\inst{9} 
\and M. Dickinson\inst{10} 
\and X. Kong\inst{11} 
\and M. Mignoli\inst{12}}

\institute{Laboratoire AIM-Paris-Saclay, CEA/DSM-CNRS--Universit\'e Paris 
Diderot, Irfu/Service d'Astrophysique, CEA Saclay, Orme des Merisiers, 
F-91191 Gif-sur-Yvette, France 
\and Institute for Astronomy, ETH Z\"urich, Wolfgang-Pauli-strasse 27, 8093 
Z\"urich, Switzerland
\and Max-Planck-Institut f\"ur extraterrestrische Physik, Giessenbachstrasse, 
85748 Garching, Germany 
\and INAF - Osservatorio Astronomico di Padova, Vicolo dell'Osservatorio 5, 
I-35122 Padova, Italy 
\and National Astronomical Observatory of Japan, Osawa 2-21-1, Mitaka, Tokyo, Japan 
\and Graduate University for Advanced Studies, Osawa 2-21-1, Mitaka, Tokyo, Japan 
\and UCO/Lick Observatory, University of California, Santa Cruz, CA 95064, USA 
\and Division of Physics, Mathematics and Astronomy, California Institute of 
Technology, Pasadena, CA 91125, USA 
\and Universit\`a di Bologna, Dipartimento di Astronomia, Via Ranzani 1, I-40127 
Bologna, Italy
\and National Optical Astronomy Observatory, P.O. Box 26732, Tucson, AZ 85726, USA 
\and Center for Astrophysics, University of Science and Technology of China, Hefei 
230026, China
\and INAF - Osservatorio Astronomico di Bologna, via Ranzani 1, I-40127 Bologna, 
Italy
}

\date{Received 08 November 2010 / Accepted 16 November 2010}

\abstract{We report evidence of a fully established galaxy cluster at $z=2.07$, 
consisting of a $\sim20\sigma$ overdensity of red, compact spheroidal galaxies 
spatially coinciding with extended X-ray emission detected with XMM-Newton.
We use VLT VIMOS and FORS2 spectra and deep Subaru, VLT and Spitzer imaging to 
estimate the redshift of the structure from a prominent $z=2.07$ spectroscopic 
redshift spike of emission-line galaxies, concordant with the accurate 12-band 
photometric redshifts of the red galaxies. Using NICMOS and Keck AO observations, 
we find that the red galaxies have elliptical morphologies and compact cores. 
While they do not form a tight red sequence, their colours are consistent with 
that of a $\gtrsim1.3$~Gyr population observed at $z\sim2.1$. 
From an X-ray luminosity of $7.2\times10^{43}$ erg s$^{-1}$ and the stellar 
mass content of the red galaxy population, we estimate a halo mass of 
5.3--8$\times10^{13}$M$_\odot$, comparable to the nearby Virgo cluster. 
These properties imply that this structure could be the most distant, mature 
cluster known to date and that X-ray luminous, elliptical-dominated clusters 
are already forming at substantially earlier epochs than previously known. 
}

\titlerunning{An old cluster in the young Universe}
\authorrunning{Gobat et al.}

\maketitle

\section{Introduction}

Massive clusters are rare structures in the distant Universe, arising from the 
gravitational collapse of the highest density peaks in the primordial spectrum 
of density fluctuations (\cite{Pee93,CL95,Pea99}). Their abundance reflects 
the original state of the matter density field and depends on fundamental 
cosmological parameters, such as the shape and normalisation of the matter 
power spectrum (e.g. \cite{PS74,Hai01,Schu03}). Measuring the distribution 
of galaxy clusters can thus place constraints on these cosmological parameters 
and provide a powerful test of primordial non-Gaussianities 
(\cite{Ji09,Ca10,Schu02}).
As the largest and most massive bound structures in the Universe, galaxy clusters 
are also the most biased environment for galaxy evolution and constitute a prime 
laboratory for studying the physical processes responsible for the formation 
and evolution of galaxies (e.g. \cite{Bo06,Par09,De10}). The strong dependence of 
galaxy activity on the surrounding environment, which gives rise to the well-known 
correlations of morphological type (\cite{Pos05,Hwa09,vdW10}) and decreasing star 
formation (\cite{Has98,Pa09,Re10,Pe10}) with increasing galaxy density, is most easily 
and dramatically illustrated in the extremely dense cores of local massive galaxy 
clusters.\\

But while, about 13.7~Gyr after the Big Bang, today's galaxies, baryons and 
dark matter continue to steadily fall into the massive clusters' potential wells, 
the elliptical galaxies that dominate their cores ceased forming stars early 
on and have been evolving passively for most of cosmological history. The traces 
of the formation process of present cluster ellipticals thus smeared out, their 
co-evolution with the cluster cores and the assembly history of the latter can hardly 
be reconstructed from low redshift data alone. Furthermore, the thermodynamical 
properties of the baryons in the intergalactic medium of clusters (ICM) suggest 
that $\sim1$~keV more energy per baryon was injected into the ICM than can be accounted 
for by pure gravitational collapse (\cite{Pon99}). This excess entropy was generated 
presumably by galactic winds powered by either supernovae or AGN, but the exact 
processes have not yet been identified. It is also unclear when the 
correlation between the mass, X-ray luminosity and temperature of the ICM, 
crucial for the use of clusters as cosmological tools, were first in place. 
The question of the assembly of clusters, the settling and thermodynamical 
evolution of the X-ray shining ICM within their deep potential wells and 
the build-up of their constituent galaxy population must then be addressed by 
looking as closely as possible at the early stages of their formation 
(\cite{Voit05,Pon99,Ro02}). This formative epoch is often put at $z\gtrsim2$, as 
supported by evidence found in recent years of increased activity 
(\cite{El07,Hay10,Hi10,Tra10}) and steeper age gradients (\cite{Ro09}) in clusters 
and overdense regions at $z\sim1-1.6$.\\

The most successful method so far for finding high-redshift galaxy clusters has 
been through X-ray searches, archival or dedicated (\cite{Ro98,Rom01,Pie03}). Their 
depth is however constrained by the sensitivity of current observing facilities, which 
limits their effectiveness at higher redshifts. Colour selection techniques, on the 
other hand, can be used to efficiently search for clusters up to $z\lesssim2$ 
(\cite{Gla00,Wi08}) and passive galaxy populations at even higher redshift (\cite{Ko07}).
Finally, galaxy clusters might also be serendipitously discovered as spatial or redshift 
overdensities. The X-ray approach naturally selects massive and evolved structures, 
while red-sequence surveys are designed to search for a distinctive evolved galaxy 
population. Indeed, $z\gtrsim1$ X-ray and colour-selected structures are spatially compact 
and dominated by massive early-type galaxies, as their local counterparts 
(\cite{Bla03,Mul05,Sta06,Pap10,Tan10,He10,Kur09}). 
In contrast, at $z>2$ the search for clusters and their precursors has focused on finding 
overdensities, often around radiogalaxies, of emission-line objects 
(\cite{Fra96,Pen97,Mil06,St05,Ov06}), in particular Ly$\alpha$ emitters. Accordingly, 
high-density structures at $z>2$ are characterised by a high level of star formation activity 
and mostly lack the extended X-ray emission and conspicuous early-type galaxy population typical 
of the evolved clusters. On the other hand, if the concentrations of Ly$\alpha$ emitters reported 
at $z>3$ (\cite{St00,Dad09,Ov08}) are proto-cluster structures destined to evolve into the massive 
X-ray clusters observed at lower redshift, we should expect to find young yet already mature 
clusters at $z\sim2-2.5$. However, whereas analogues to local massive clusters are known up to 
$z=1.5$--1.7, evolved early-type dominated and X-ray emitting clusters have not been found so 
far at earlier epochs, nor has an ``intermediate'' structure been observed at (or right after) 
the moment of quenching of star formation in its core galaxies.\\

Here we present the discovery of CL J1449+0856, a conspicuous galaxy overdensity at $z=2.07$, 
dominated by massive passively evolving galaxies and consistent with being the most distant 
X-ray detected galaxy cluster identified to date.
In Section \ref{phot}, we describe the target selection and photometric observations. 
In Section \ref{morph}, we present the high resolution imaging and morphological 
analysis of galaxies in the core and in Section \ref{spec} we discuss the spectroscopic 
observations and redshift confirmation of the cluster. In Section \ref{X}, we discuss the 
X-ray observations of the cluster and their analysis, while in Section \ref{proto} we 
compare CL J1449+0856 to other high-redshift structures. In Section \ref{disc}, we discuss its 
global properties and their implications for cosmology and Section \ref{sum} summarises our 
results. Unless specified otherwise, all magnitudes are reported in the AB system and we adopt 
a concordance cosmology with $H=70$~km~s$^{-1}$~Mpc$^{-1},\Omega_m=0.3$ and $\Lambda=0.7$.

\section{\label{phot}Imaging and sample selection}

The structure was first identified, in archival Spitzer images covering 342~arcmin$^2$ of the 
so-called ``Daddi Field'' (\cite{Dad00}), as a remarkable overdensity of galaxies with IRAC 
colours $[3.6]-[4.5]>0$ at the position RA = 14h 49m 14s and DEC = 8$^{\circ}$ 56' 21", indicating 
a massive structure at $z>1.5$. Optical-NIR imaging data of this field was already available from 
a multi-band survey in the $B,R,I,z,$ and $K_s$ bands, the latter somewhat shallow. These data 
and their reduction are described in \cite{Kong06} and \cite{Dad00}.
Between 2007 and 2010, we obtained new deep imaging of this overdensity in the $Y,J,H,$ 
and $K_s$ bands with MOIRCS on the Subaru telescope and in the $J$ and $K_s$ bands with 
ISAAC on the VLT. For the purpose of studying galaxy morphology, we also obtained deep $F160W$ 
imaging of the overdensity with NIC3 on Hubble and a shallow but high resolution $K$-band image 
using NIRC2 with adaptive optics on Keck. These two images were not used in the making of the 
photometric catalogue and their analysis is described in Section \ref{morph}. In addition to the 
archival Spitzer/IRAC images, archival 24~$\mu$m data taken with Spitzer/MIPS were also available. 
We mention them for completeness but discuss them only briefly here, as they will be included in 
a future analysis of the galaxy population of the structure. Finally, X-ray observations of the 
field were also available and are described in Section \ref{X}. Details of the imaging observations 
are given in Table \ref{tab:phot}.

The combined $B,R,I,z,Y,J,H,$ and $K_s$ observations reach 5$\sigma$ limiting magnitudes of 26.95, 
26.18, 26.03, 25.81, 25.64, 25.47, 23.66 and 24.74 respectively. Catalogues were made for each band with 
SExtractor (\cite{BA96}) using 2''-diameter apertures and later merged. The final catalogue covers 
an area of $4'\times7'$, corresponding to the field of view of MOIRCS.
As the galaxies with $[3.6]-[4.5]>0$ are better detected in the $Y$-band image than in the $z$ and 
$B$-band images, and since $Y$ straddles the 4000 $\AA$ break at $z=1.5$, we do not rely on the 
traditional $BzK$ criterion (\cite{Dad04}) to select for passively evolving galaxies at $z>1.5$ but 
instead use $Y-K_s>2$, the expected colour of such a galaxy population. Out of 1291 objects in the 
combined catalogue with $K_s<24.74$, we find 114 red galaxies with $Y-K_s>2$. We note that only 11 
of those are detected at 5$\sigma$ in $B$ and 41 in $z$.\\

\begin{figure*}
\centering
\includegraphics[width=0.49\textwidth]{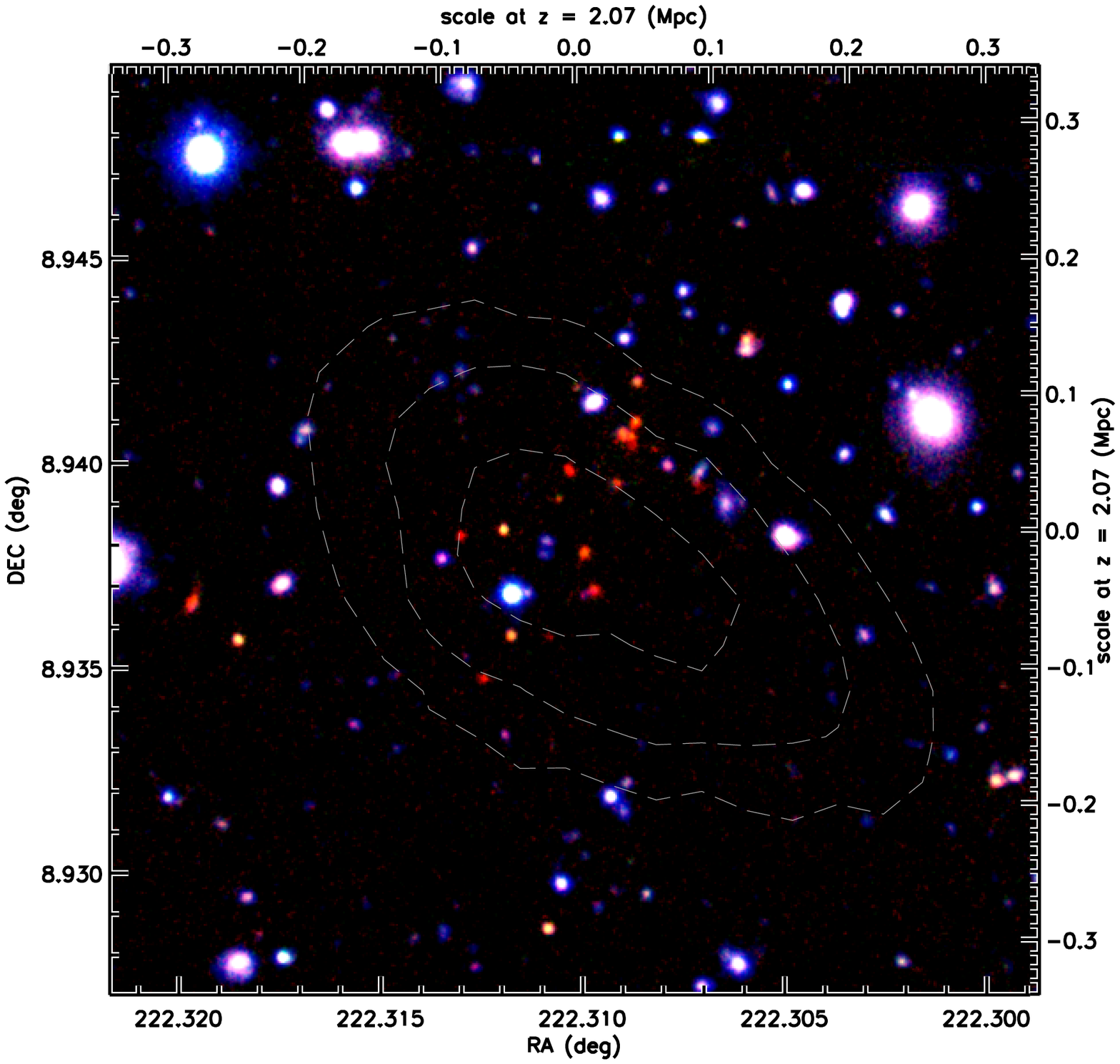}
\includegraphics[width=0.49\textwidth]{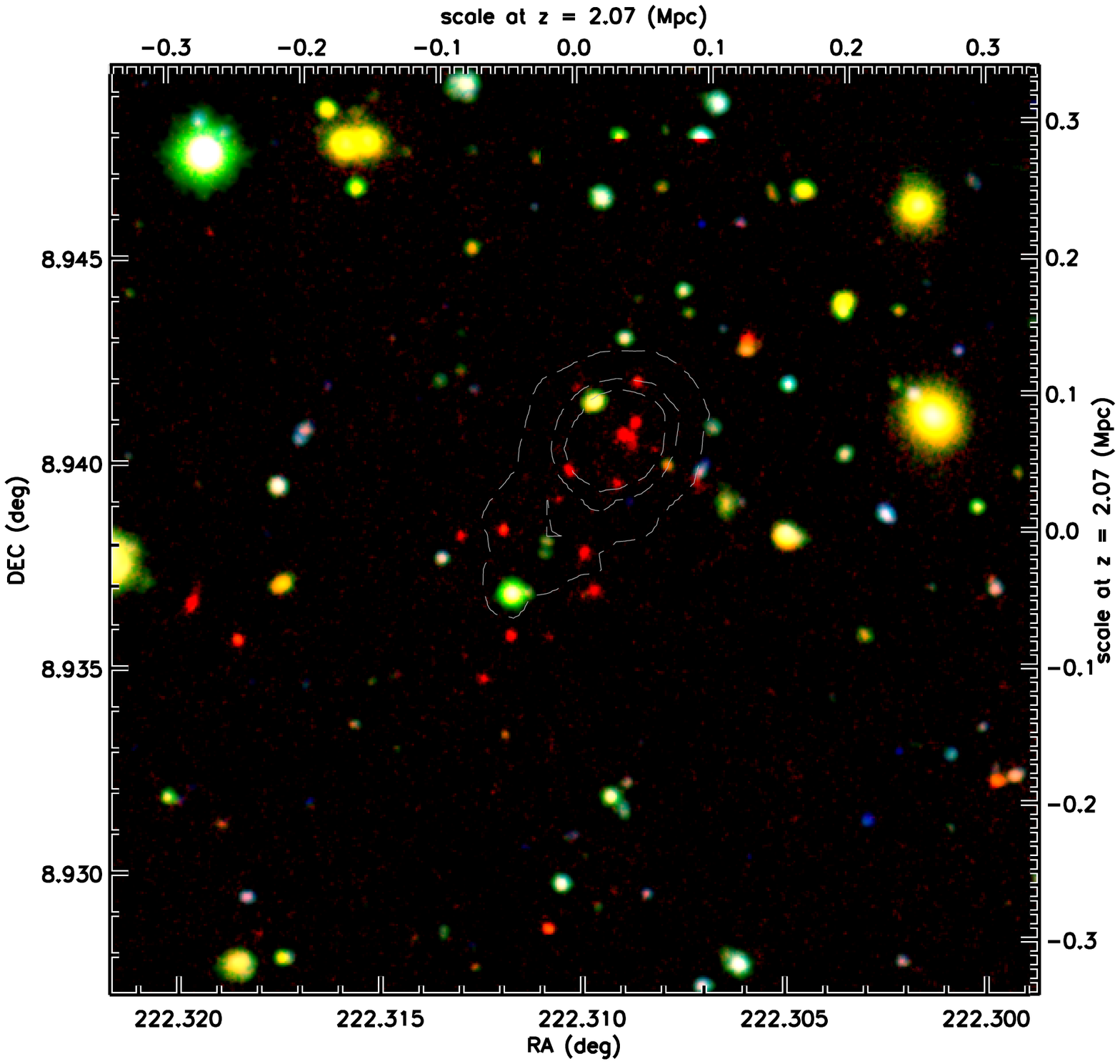}
\caption{RGB composite colour images of the 1.4'$\times$1.4' field centred on the 
galaxy overdensity. The R channel of both images corresponds to the $K_s$ band; the G 
and B channels corresponds to the $J$ and $z$ bands in the left image and to the 
$B$ and $z$ bands in the right one. The $B$ and $z$ images were taken using the 
Suprime-Cam instrument on the Subaru telescope while the $J$ and $K_s$ images are 
a composite of MOIRCS and ISAAC data, on the Subaru and VLT observatories. The $BzK_s$ 
image shows how the red galaxies are basically unseen at optical wavelengths.
On the left, the white overlapping contours show the 1, 2, and 3$\sigma$ significance 
levels of the diffuse X-ray emission in the XMM-Newton image, after subtraction of a 
point source seen in the Chandra image and smoothing with a 8'' radius PSF (as described 
in Section \ref{X}). On the right, they show the 10, 20, and 30$\sigma$ levels above the 
background galaxy number density, computed using the $\Sigma_5$ estimator.}
\label{fig:rgb1}
\end{figure*}

\begin{figure*}
\centering
\includegraphics[width=0.8\textwidth]{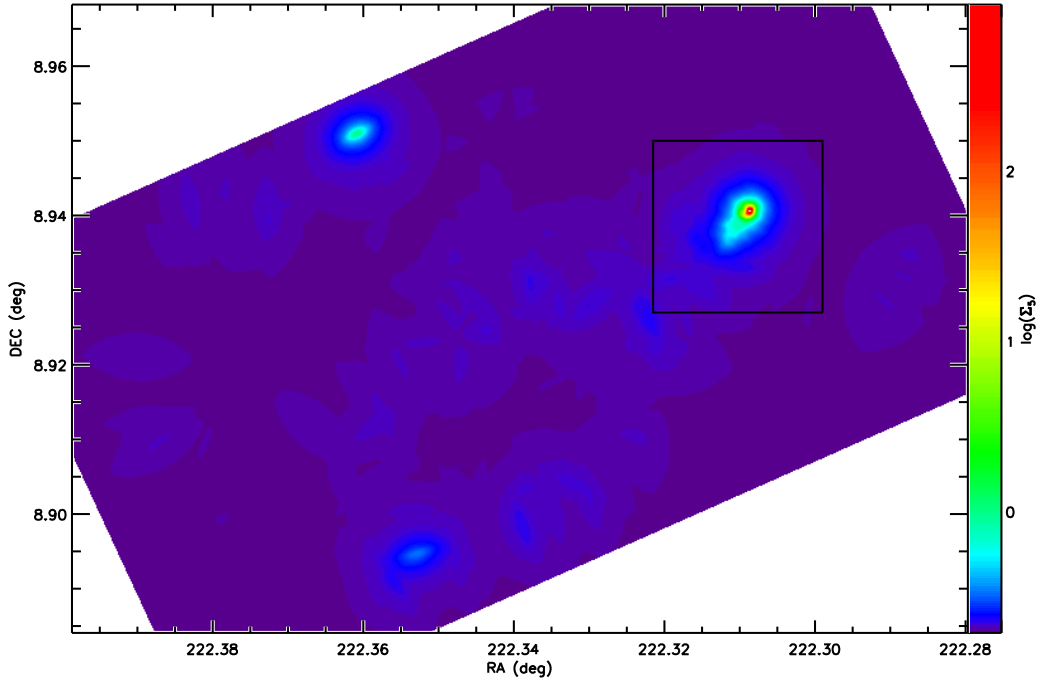}
\caption{Surface density of galaxies with $Y-K_s>2$ in the whole MOIRCS field, in 
units of galaxies arcmin$^{-2}$ as measured by the $\Sigma_5$ estimator. The black box 
delimits the 1.4'$\times$1.4' field shown in  Fig. \ref{fig:rgb1}.}
\label{fig:dens}
\end{figure*}

The distribution of red galaxies shows a strong overdensity at the same position as the IRAC-selected 
one. To characterise the overdensity, we created a number density map by dividing the MOIRCS field 
into a grid of sub-arcsecond cells and computing for each element the density estimator 
$\Sigma_N\equiv N/\pi r_N^2$, where $r_N$ is the distance to the Nth nearest galaxy with $Y-K_s>2$. 
We considered $N=3-7$, which changed the angular resolution of the map but produced consistent 
results. The MOIRCS detector consists of two chips, one of which was centred on the overdensity and 
the other thus providing a low-density field. In the near-infrared images, the visible overdensity 
defines a 20''$\times$10'' semi-axis elliptical area. We find that, in this 20'' region centred on the 
structure, the mean density is $20\sigma$ above the field, with $\sim100$ galaxies arcmin$^{-2}$ in the 
overdensity versus five in the field. The density of this structure is thus similar to that reported for 
the recently discovered galaxy cluster at $z=1.62$ (\cite{Pap10}).
Fig. \ref{fig:rgb1} shows two colour images of the overdensity, with contours representing X-ray 
intensity and galaxy density, respectively, while Fig. \ref{fig:dens} shows the number density of red 
galaxies in the MOIRCS field.\\

\begin{table*}
\centering
\scriptsize
\begin{tabular}{ccccccc}
Filter&Central wavelength ($\mu$m)&Exposure (s)&5$\sigma$ limit (mag)&Instrument&Telescope&Observation date\\
\hline
$B$&0.44&1500&26.95&Suprime-Cam&Subaru&2003 Mar 5\\
$R$&0.65&3600&26.18&Suprime-Cam&WHT&1998 May 19-21\\
$I$&0.80&1800&26.03&Suprime-Cam&Subaru&2003 Mar 5\\
$z$&0.91&2610&25.81&Suprime-Cam&Subaru&2003 Mar 4-5\\
$Y$&1.02&17780&25.64&MOIRCS&Subaru&2009 Mar 15, 2010 Feb 7-8, 21\\
$J$&1.26&9360&25.47&MOIRCS+ISAAC&Subaru+VLT&2007 Mar 10, Apr 5\\
$F160W$&1.60&17920&&NIC3&HST&2008 May 11\\
$H$&1.65&2380&23.66&MOIRCS&Subaru&2007 Apr 8\\
$K$&2.15&1890&&NIRC2&Keck&2009 Apr 4\\
$K_s$&2.20&7800&24.74&MOIRCS+ISAAC&Subaru+VLT&2007 Mar 8, Apr 5\\
\hline
IRAC 1&3.6&480&23.85&IRAC&Spitzer&2004 Jul 22\\
IRAC 2&4.5&480&23.08&IRAC&Spitzer&2004 Jul 22\\
IRAC 3&5.8&480&21.44&IRAC&Spitzer&2004 Jul 22\\
IRAC 4&8.0&480&20.02&IRAC&Spitzer&2004 Jul 22\\
MIPS 24&24&480&80~$\mu$Jy&MIPS&Spitzer&2004 Aug 5\\
\hline
0.5-10 keV&&80000&&EPIC-MOS&XMM-Newton&2001-2003\\
0.5-8 keV&&80000&&ACIS&Chandra&2004 Jun 7-13\\
\end{tabular}
\small
\caption{Details of the optical, near-IR photometric and X-ray observations. 
The $B$, $R$, $I$ and $z$-band data have already been described in a previous 
paper (\cite{Kong06}). The XMM and Chandra data have also been described in other 
papers (\cite{Bru05,Cam09}). The raw IRAC and MIPS data were taken from the 
archive and reduced using MOPEX and custom scripts.}
\label{tab:phot}
\end{table*}

\section{\label{morph}High resolution imaging and galaxy morphology}

Very dusty star-forming galaxies at high redshift can have red colours 
similar to those expected for elliptical galaxies. While a complete morphological study of 
galaxies in the overdensity will be the subject of a future publication, we carried out 
here a first-order analysis to assess the nature of the red galaxies. We used $H$-band 
observations carried out with NICMOS-3 on Hubble, which provides the required sensitivity 
and spatial resolution to unveil the morphology of the red galaxies, as well as 
AO-assisted ground-based imaging with NIRC2 on Keck. The NIC3 data were taken during seven 
orbits of Hubble. The individual frames were reduced using the NICRED pipeline (\cite{Mag07}), 
which we found provides a better calibration than the standard pipeline. In particular, 
the background noise varies less in the NICRED-reduced images, which is critical to 
the morphological analysis as it allows us to recover the extended luminosity profile 
of faint galaxies. The individual frames were then combined using \emph{Multidrizzle} 
(\cite{Koe02}). The useful area of the NIC3 image, where the noise is low enough for the 
morphological analysis, is 45''$\times$50'', covering the overdensity and its immediate 
surroundings. 

\begin{figure*}
\centering
\includegraphics[height=0.8\textwidth,angle=90]{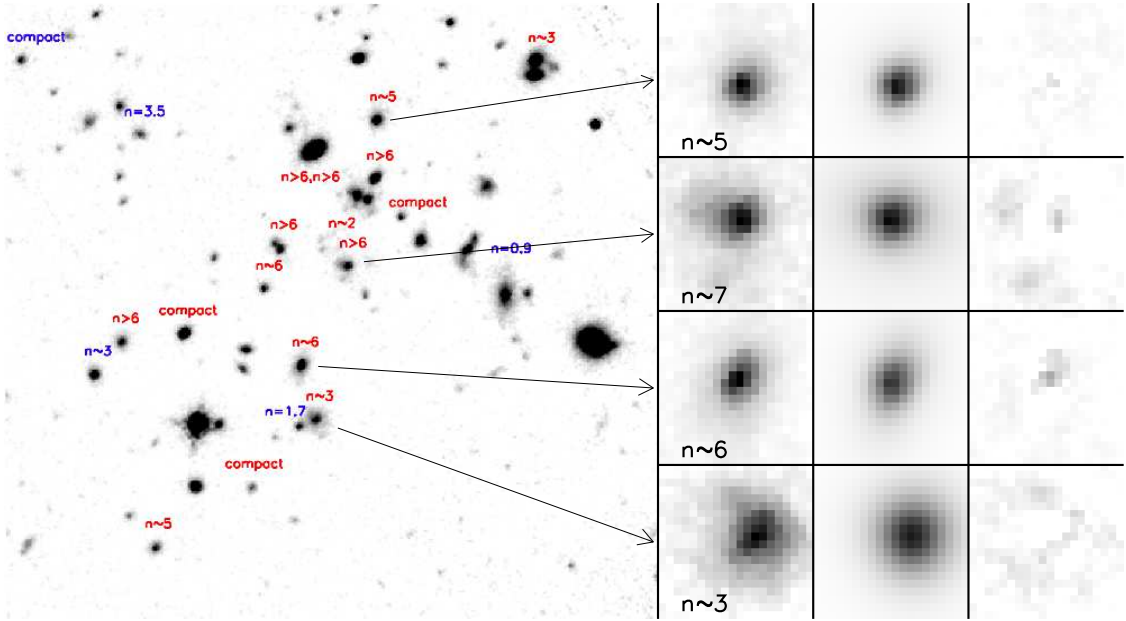}
\caption{Morphological properties of the galaxies in the field of the overdensity. 
Left, $H$ band ($F160W$) NICMOS image of the cluster, in logarithmic greyscale. 
The image is a composite of frames taken with the NIC3 camera during 7 HST orbits, 
reduced using the NICRED pipeline and combined with Multidrizzle. Galaxies with 
$Y-K_s>2$ are shown in red and galaxies with $Y-K_s<2$ in blue. S\'ersic indices 
of the best-fit model are indicated on top of each galaxy, except for those that 
are too point-like, which are labelled ``compact''. Right, morphologies of four 
representative galaxies from the $H$ band image. For each galaxy we show, from left 
to right, the observed image, the best-fit S\'ersic model and the residuals image 
after subtraction of the model, all in the same logarithmic grey scale.}
\label{fig:morph1}
\end{figure*}

The galaxies in this image were first matched with the catalogue and their surface 
brightness profiles modelled with a S\'ersic law ($\propto r^{1/n}$; \cite{Ser63}) using 
GALFIT (\cite{Peng02}) and the single star in the NIC3 field as PSF. We fitted both the red 
($Y-K_s>2$) and blue ($Y-K_s<2$) galaxies. For the latter, we used the $BzK$ criterion to 
select for star-forming galaxies at $z>1.4$: there are 352 $sBzK$-selected galaxies in the 
catalogue detected at 5$\sigma$ in all three bands, of which ten are found in the NIC3 image. 
In that same field, we find 16 red galaxies. Fig. \ref{fig:morph1} shows the NIC3 image and 
the results of the S\'ersic fit: of those 16 red galaxies, five are unresolved (i.e. have an 
effective radius of one pixel or less; labelled ``compact'' in Fig. \ref{fig:morph1}) and 
the rest have $n>4$, whereas three of the ten blue galaxies are unresolved, two have $n\gtrsim3$, 
and the rest $n\lesssim2$.\\

\begin{figure*}
\centering
\includegraphics[width=0.8\textwidth]{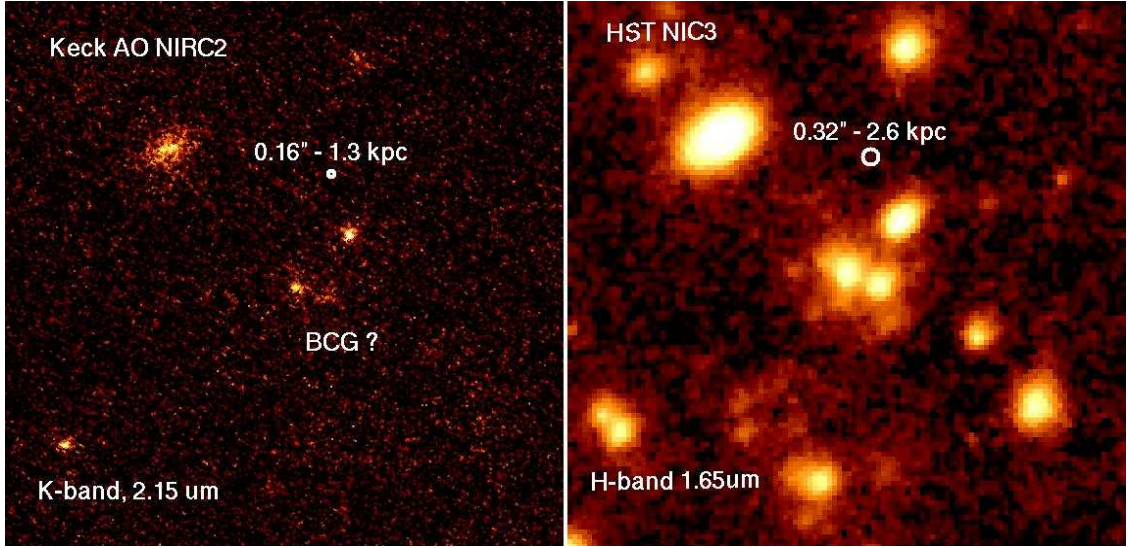}
\caption{Comparison of space and ground-based near-IR images of the field around the 
seemingly interacting galaxy triplet, taken with the NIC3 instrument on the HST (right) 
and NIRC2 on Keck using adaptive optics (left) and showing the compact cores of two of 
the three galaxies. In both images, the PSF is shown by a white circle.}
\label{fig:morph2}
\end{figure*}

The NIRC2 image covers a similar area to the NIC3 image, but is rather shallow, having been 
taken during an unrelated observation, and only the brightest galaxy cores are visible.
It however provides a good first-order verification of the S\'ersic modelling. Because of its 
combination of shallowness and very high resolution, the objects visible in the NIRC2 image 
either would be intrinsically bright or, in the case of faint galaxies like those in the 
overdensity, have a high surface brightness, similar to that of compact cores. 
Fig. \ref{fig:morph2} compares the NIRC2 and NIC3 images. We find that, 
of the 13 red galaxies in the useful area of the NIRC2 image, eight are distinctly visible as 
very compact sources, supporting the results of the S\'ersic modelling. None of the $sBzK$ 
galaxies rises above the noise of the NIRC2 image. The rest of the visible objects 
are obvious interlopers, low redshift galaxies (colour- and angular size-wise) and a star. 

The galaxy overdensity is thus clearly dominated by spheroidal galaxies, as in local 
massive clusters. We note that more than half of these galaxies have $r_e>2$~kpc and thus 
appear less dense than previously studied passive galaxies at $z>2$ (\cite{Tof07}). A 
thorough discussion of the stellar mass-size relation will be presented in a future paper.\\

\section{\label{spec}Spectroscopic observations and redshift determination}

\begin{figure*}
\centering
\includegraphics[width=0.8\textwidth]{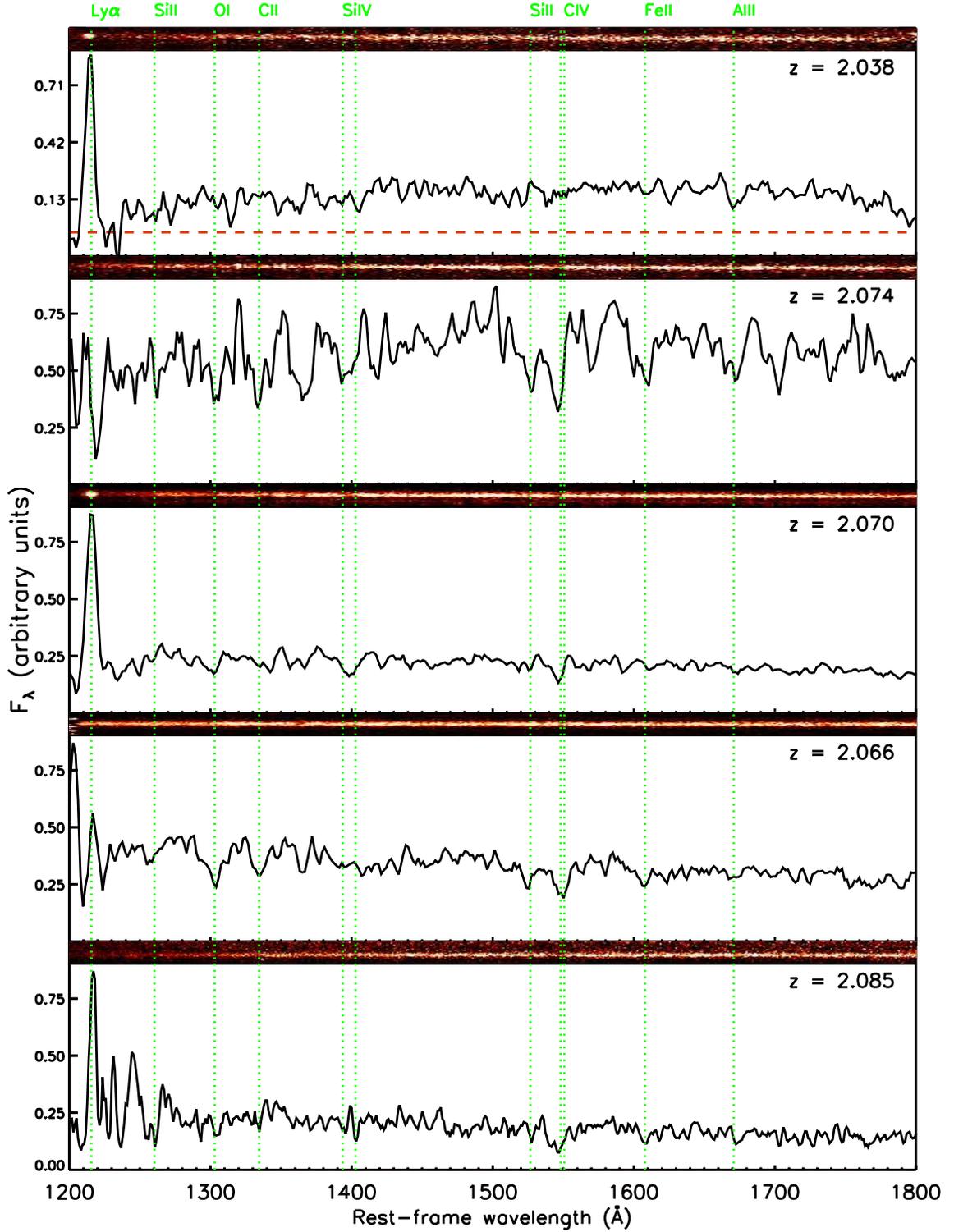}
\caption{Rest-frame UV spectra, taken with FORS2 and VIMOS on the VLT, of blue 
star-forming galaxies with redshifts within the peak of the distribution, centred 
at $z=2.07$. The spectra shown here were rebinned with a bin size of two pixels. 
Prominent emission and absorption features are shown with dotted green lines. 
The dashed red line indicates the level of zero flux. For each object, the (uncalibrated) 
image from which the spectrum was extracted is shown at the top of the sub-plot.}
\label{fig:spec}
\end{figure*}

\begin{figure*}
\centering
\includegraphics[width=0.49\textwidth]{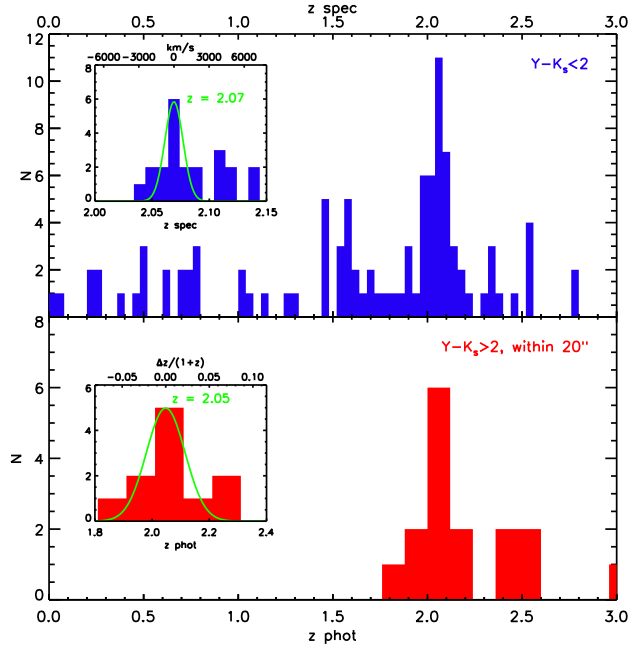}
\caption{Redshift distributions of the galaxies in and surrounding the overdensity. 
Top: distribution of redshifts, determined from the emission-line spectra of blue 
($Y-K_s<2$) galaxies taken with the FORS2 and VIMOS instruments on the Very Large Telescope. 
The insert shows a more detailed histogram for the range $z=2-2.15$. For galaxies 
in the peak, the spectra sample the rest-frame $\sim1200-2200$\AA. Bottom: distribution 
of photometric redshifts of red ($Y-K_s>2$) galaxies within the central 20''. These were 
estimated by comparing the $BIRzYJHK_s$+IRAC spectral energy distribution to a set of 
spectral synthesis models computed from templates of half-solar, solar, and twice-solar 
metallicity and assuming two different types of star-formation history (exponentially 
declining and constant truncated). We did not include the effects of dust attenuation, 
as the colour, morphological, and spectral properties of the red galaxies point strongly 
towards them being dominated by passive stellar populations rather than dust-reddened 
younger stars.}
\label{fig:z}
\end{figure*}

We performed spectroscopic follow-up observations of the galaxies in and around the 
overdensity using FORS2 and VIMOS on the VLT and MOIRCS on Subaru.
Blue, $sBzK$-selected galaxies were targeted around the structure's centre to within 10' 
(or 5~Mpc at $z\sim2$; see Fig. \ref{fig:map1}). The VLT observations consisted of two 
FORS2 masks with 5 hours of integration each and one VIMOS mask with a 2.5h exposure. 
The 2D spectra of the individual runs were reduced using the standard pipeline (\cite{Sco05})
and co-added. One-dimensional spectra were then extracted using the \emph{apextract} tasks of 
the IRAF package. We estimated redshifts by first cross-correlating the observed spectra with 
a set of templates, including Lyman-break (\cite{Shap03}), starburst, and star-forming 
(\cite{Kin96}) galaxies. Using the rough redshift estimates given by the peaks of the 
cross-correlation function, we derived more precise redshifts from emission and absorption 
features (at $z>1.4$, the FORS2 and VIMOS spectra cover the rest-frame UV) using the 
\emph{rvidlines} task of the IRAF package. From the 41 FORS2 and 164 VIMOS slit spectra taken, 
we determined 109 secure redshifts. Their distribution shows a clear spike in the range 
$z=2-2.1$, as shown in the top panel of Fig. \ref{fig:z}. Assuming that this distribution 
is Gaussian, it peaks at $z=2.07$ and has a dispersion of $\sim780\pm90$~km/s. Using the 
biweight estimator (\cite{Be90}), we find $z=2.07$ and 747~km/s respectively. These values 
of cluster velocity dispersion are comparable to that of star-forming galaxies in the nearby 
Virgo cluster (\cite{Bing87}). With these values, we find 11 galaxies having spectroscopic 
redshifts within 2$\sigma$ of the $z=2.07$ peak. Some representative spectra of galaxies 
within the redshift spike are shown in Fig. \ref{fig:spec}.\\

We also obtained spectra of several red galaxies in the near-IR with OHS/CISCO and MOIRCS 
on Subaru, but their faintness prevented us from measuring redshifts: 
while a stacked spectrum of the brightest member candidates shows some continuum, no 
absorption or emission features are seen, the latter down to typical limits of 
$\sim7\times10^{-17}$~erg~s$^{-1}$~cm$^{-2}$. At $z\sim2$, this value corresponds to an 
unreddened star formation rate of $<10$~M$_{\odot}$~yr$^{-1}$ and tends to support the 
conclusion that these objects are passively evolving stellar populations.\\

\begin{figure*}
\centering
\includegraphics[width=0.49\textwidth]{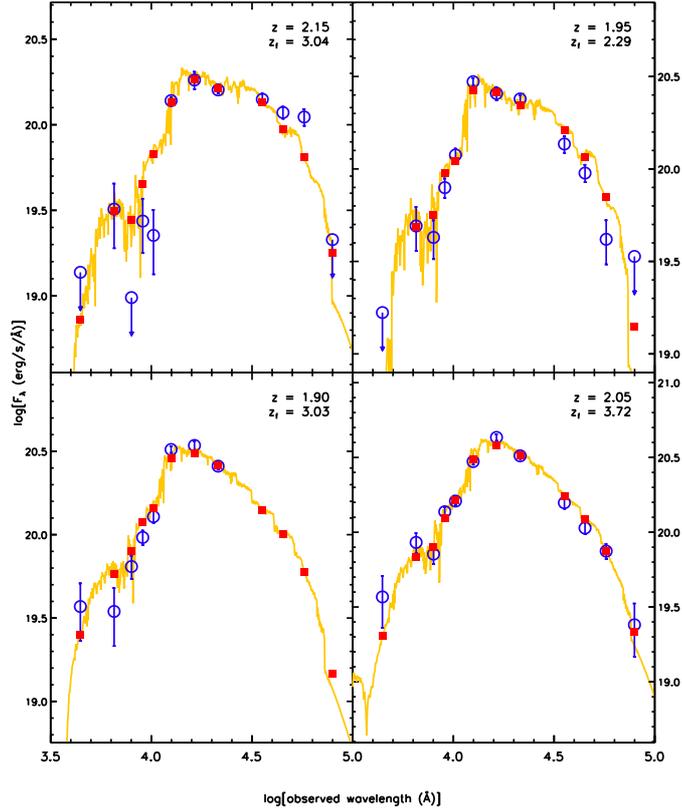}
\caption{Spectral energy distributions of four representative red galaxies with 
photometric redshifts within $2\sigma$ of the peak of the distribution. The observed 
SEDs and errors are represented with blue open circles. Upper limits at $1\sigma$ are 
shown by arrows and the best fit template is shown in orange. 
The corresponding redshift and formation redshift, at which half of the stellar mass 
was in place, are given in the upper right corner of each sub-plot. The integrated 
template fluxes in the Suprime-Cam, MOIRCS and IRAC bands are shown by red squares.}
\label{fig:seds}
\end{figure*}

\begin{figure*}
\centering
\includegraphics[width=0.49\textwidth]{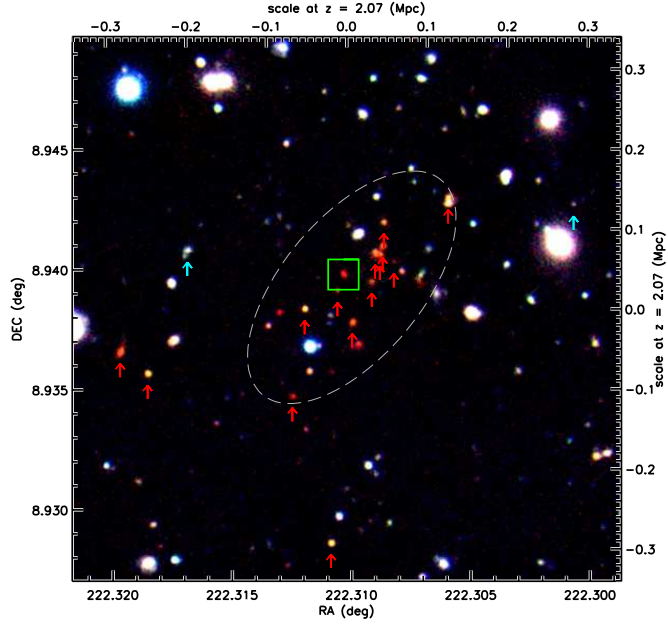}
\caption{RGB composite image of the 1.4'$\times$1.4' field centred on the CL J1449+0856, as 
in Fig. \ref{fig:rgb1}. 
The R, G, and B channels correspond to the $K_s, J,$ and $Y$ bands, respectively. 
The $Y$, $J,$ and $K_s$ photometry straddles wavelengths close to the 4000\AA\ break at $z\sim2$ 
and is useful to appreciate likely age variations among potential cluster members. 
The red arrows show the positions of red galaxies with photometric redshifts within 2$\sigma$ 
of the peak of the distribution and the blue ones the position of galaxies with spectroscopic 
redshifts within 2$\sigma$ of the peak. The green square gives the position of the AGN seen 
with Chandra. The white ellipse shows the 20'' semi-major axis region that corresponds roughly to 
the overdensity and from which were selected the red galaxies shown in Figs. \ref{fig:seds} 
and \ref{fig:cmd} and in Table \ref{tab:zphot}.}
\label{fig:rgb2}
\end{figure*}

\begin{figure*}
\centering
\includegraphics[width=0.49\textwidth]{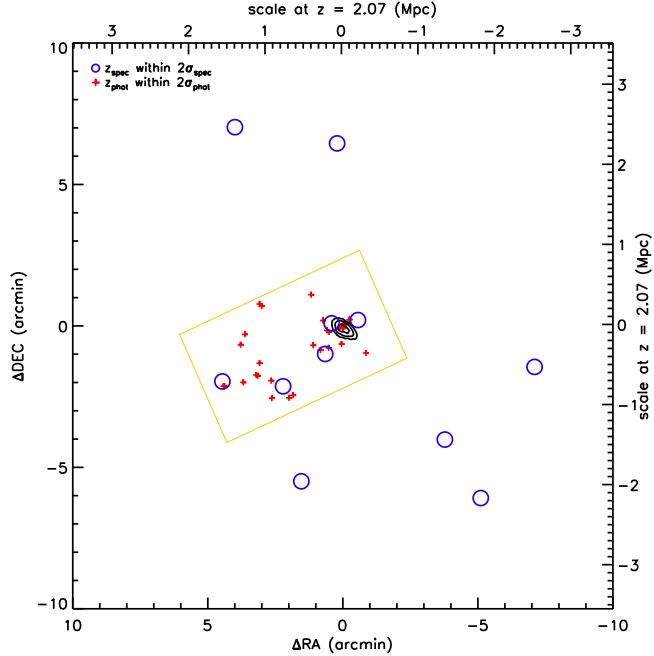}
\caption{Spatial distribution of blue spectroscopic members, centred on the cluster. Galaxies 
with spectroscopic redshifts within $2\sigma$ of the $z=2.07$ peak are shown by blue open 
circles. The positions of galaxies with $z_{phot}$ within $2\sigma$ of the $z=2.05$ peak are 
shown by red crosses. The orange rectangle shows the field of MOIRCS and the 1, 2, and 3$\sigma$ 
significance levels of the extended X-ray emission (as shown on Fig. \ref{fig:rgb1} and discussed 
in Section \ref{X}) are drawn at the centre of the plot.}
\label{fig:map1}
\end{figure*}

To complement the spectroscopic redshift information, which does not include 
the red galaxies, we estimated photometric redshifts from our $BIRzYJHK_s$+IRAC 
photometry. Some galaxies in the catalogue have $[5.8]-[8.0]>0$, suggesting that 
emission from an obscured AGN is contributing to the infrared SED and that the latter can 
therefore not be reproduced well by standard stellar population models. In particular, 
the galaxy in the overdensity with the highest IRAC excess is detected in the X-ray data, 
as discussed in Section \ref{X}. In these cases, we ignored the IRAC data for the purpose of 
deriving photometric redshifts.
We compared the SEDs of the red galaxies to a range of model SEDs obtained 
from a set of \cite{M05} stellar population synthesis templates, with three different 
star-formation histories (single burst, exponentially declining, and a constant star formation 
rate with truncation) and metallicities ranging from half to twice the solar value. We did not 
include the effects of dust extinction in the models, as the colour and morphology of the red 
galaxies, as well as the absence of emission lines in the spectra of those targeted with MOIRCS 
and OHS/CISCO, strongly suggest that they form a population of passively evolving systems. 
Fitting the SEDs with actively star-forming stellar populations and an arbitrary amount 
of dust results in a substantially worse $\chi^2$.
We also used the SED fit to estimate ages and stellar masses, assuming a bottom-light initial 
mass function (\cite{Cha03}). Fig. \ref{fig:seds} shows the SEDs and best-fit models of four 
red galaxies in the structure. 
We find that the distribution of the photometric redshifts of red galaxies in the 
overdensity proper (i.e. within 20'' from the structure's centre; see Fig. \ref{fig:rgb2}) 
narrowly peaks at $z=2.05$ (Fig. \ref{fig:z}), with a scatter of $\sim0.07$ that is 
fully compatible with the expected accuracy for high-redshift ellipticals (\cite{Dad05,Il06,M06}). 
We take the narrowness of both redshift distributions and their very similar peak values 
as confirmation that this structure, hereafter CL J1449+0856, is a real cluster or proto-cluster 
and we thus set its most likely redshift as $z=2.07$. 
For the rest of the analysis, we consider as members of the structure galaxies with spectroscopic 
or photometric redshifts within 2$\sigma$ from the peaks of their respective distributions. 
We note that while $z=2.07$ is the most likely redshift for this structure, having not 
yet measured any spectroscopic redshift for the red ellipticals and based on the widths 
of the spectroscopic and photometric redshift distributions, we cannot exclude a slightly 
different redshift in the range $2\lesssim z\lesssim2.1$.

Fig. \ref{fig:map1} shows the spatial distribution of spectroscopic and photometric 
members compared to the size of the core and the field of view of MOIRCS. Table \ref{tab:zphot} 
gives the characteristics of the photometric members in the core of CL J1449+0856.

\begin{table*}
\centering
\small
\begin{tabular}{ccccccccc}
ID&RA (J2000)&DEC (J2000)&$z_{phot}$&$Y-K_s$ (AB)&$K_{s,tot}$ (AB)&$r$ (')&$r$ (kpc)\\
\hline
4774&14:49:14.39&+8:56:16.19&2.15&3.73$\pm$0.44&22.33&0.09&43\\
4942&14:49:14.20&+8:56:22.25&2.10&2.82$\pm$0.38&22.73&0.05&27\\
4948&14:49:14.54&+8:56:20.86&2.10&2.15$\pm$0.83&23.85&0.04&18\\
4954&14:49:14.87&+8:56:18.14&1.95&2.36$\pm$0.10&22.15&0.13&65\\
4970&14:49:13.97&+8:56:25.20&1.95&2.34$\pm$0.50&23.68&0.12&62\\
5080&14:49:14.11&+8:56:26.30&1.90&3.13$\pm$0.26&21.76&0.11&54\\
5125&14:49:14.08&+8:56:27.59&2.05&3.08$\pm$0.18&21.91&0.13&65\\
5138&14:49:14.16&+8:56:26.53&2.25&2.70$\pm$0.13&21.58&0.10&52\\
5225&14:49:13.42&+8:56:34.81&2.05&2.36$\pm$0.09&21.40&0.33&165\\
5262&14:49:14.07&+8:56:31.19&2.10&2.31$\pm$0.18&22.55&0.18&91\\
\hline
4489&14:49:14.99&+8:56:05.06&2.25&2.77$\pm$1.10&23.04&0.31&155\\
4518&14:49:14.60&+8:56:08.75&2.45&2.42$\pm$0.48&24.00&0.22&108\\
4608&14:49:14.33&+8:56:12.87&2.50&2.30$\pm$0.15&22.22&0.14&72\\
5018&14:49:14.10&+8:56:25.50&4.70&2.68$\pm$0.55&22.55&0.10&51\\
5045&14:49:15.13&+8:56:17.63&2.75&3.04$\pm$0.36&23.01&0.19&96\\
5113&14:49:14.48&+8:56:23.37&3.05&3.82$\pm$0.45&22.47&0.04&19\\
\end{tabular}
\caption{Characteristics of the red galaxies in the overdensity: coordinates, photometric 
redshift, $Y-K_s$ colour, $K_s$-band total magnitude, and distance 
from the cluster centre in arc minutes and kiloparsecs. The first ten galaxies 
are the red members, with $1.9<z_{phot}<2.2$. The bottom six objects, which are 
not formally inside the photometric redshift peak, might still belong to the structure: 
e.g. galaxy 5113, which is the Chandra-detected AGN, a red spheroidal ($n_{Sersic}=6$) 
whose optical-NIR SED likely suffers from AGN contamination.}
\label{tab:zphot}
\end{table*}

\subsection{Properties of the red galaxy population}

\begin{figure*}
\centering
\includegraphics[width=0.49\textwidth]{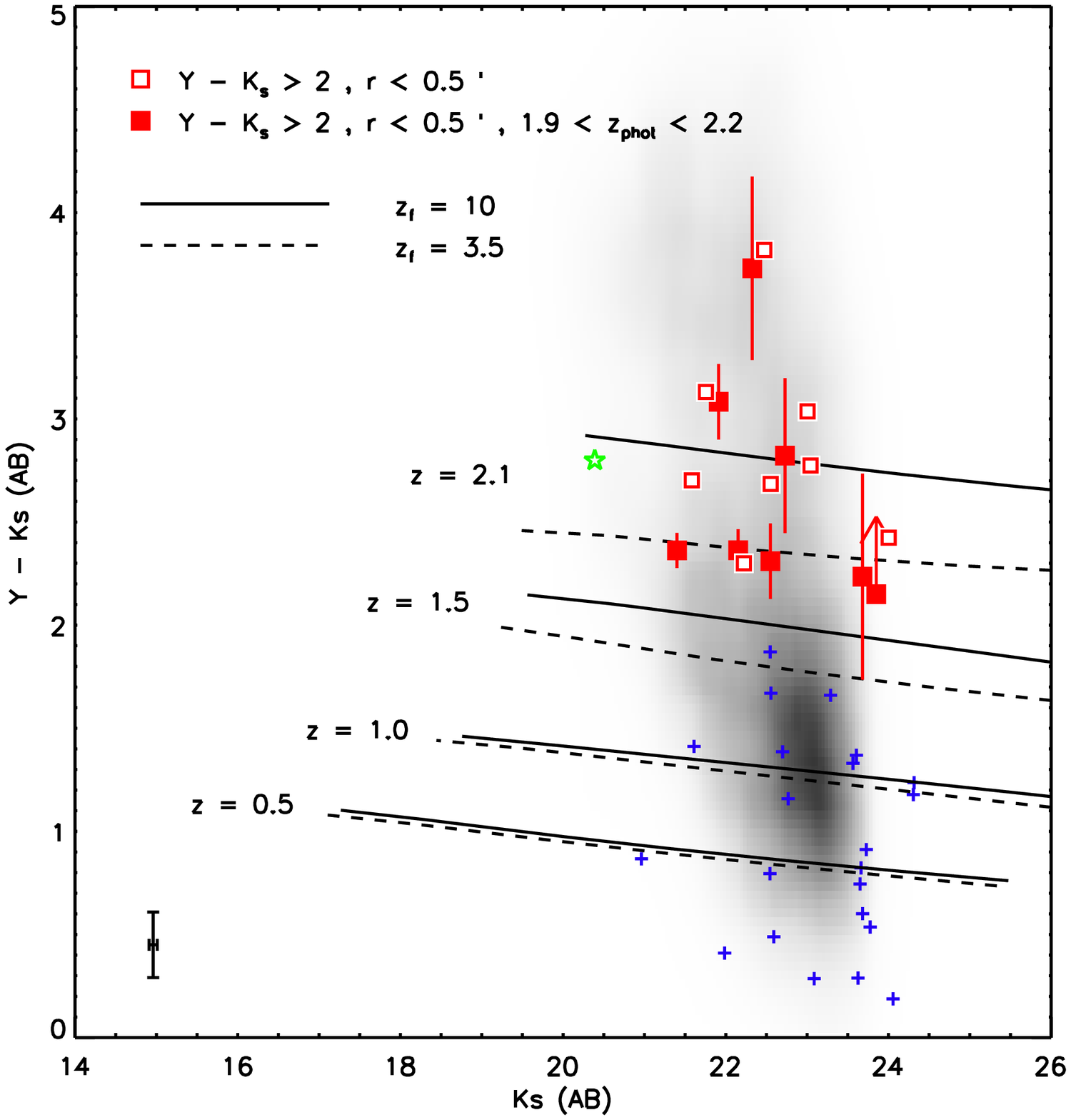}
\caption{Colour-magnitude diagram of the objects within 20" of the cluster 
centre. Blue galaxies are shown by blue crosses and red galaxies by red squares. 
Red galaxies with a photometric redshift within 2$\sigma$ of the peak of the 
distribution are indicated by filled squares, the bars showing the error in the 
colour. The greyscale shaded map shows the distribution of $BzK$-selected galaxies in 
the GOODS-South field (\cite{Dad07a}).
Colour limits at $3\sigma$ are shown by arrows. The typical error in the 
$Y-K_s$ colour and $K_{s,tot}$ magnitude is shown in the bottom left. The solid 
and dashed lines show the expected colour and magnitude of stellar population 
synthesis models (\cite{KA97}), assuming that star formation begins at $z=10$ and 
$z=3.5$, respectively. The composite flux and colour of the ``proto-BCG'' galaxy 
assemblage are marked by a green star.}
\label{fig:cmd}
\end{figure*}

The colours of those red photometric members are significantly redder than those 
observed for established elliptical galaxies in $z\sim1.5$ clusters (e.g. \cite{Pap10}) and 
are consistent with a passive population at $z\sim2.1$, as shown in Fig. \ref{fig:cmd}. We 
find a large colour scatter, however, indicating that the ellipticals have not yet 
settled into a tight red sequence characteristic of $z\lesssim1.6$ clusters. 
Indeed, their star-formation weighted ages (the mean age of the stars in the best-fit 
model) range from 0.6 to 2 Gyr, with an average of 1.2 Gyr and average 
formation redshift of $z\sim3.5$. This implies that, if truly passive, some of these 
galaxies are observed relatively shortly after the cessation of star formation 
and before the colour differences due to their different star-formation histories could 
be attenuated by the subsequent passive evolution.\\

None of the elliptical members is much brighter than the others as  to qualify for 
being a ``brightest cluster galaxy'' (BCG), the brightest of the red members being 
less than 0.5 mag more luminous than the second and third brightest members. However, 
as seen in Figs. \ref{fig:morph1} and \ref{fig:morph2}, a group of three galaxies near 
the centre of the cluster appear to be very close to each other. If they all are at 
the same redshift of $z=2.07$, they are separated by 5.5 to 13 kpc in projection and thus 
likely interacting. As noted in Section \ref{morph}, Keck adaptive optics observations 
with NIRC2 reveal in two of them a very compact ($\sim1$~kpc) core typical of 
high-redshift elliptical galaxies (\cite{Dad05}). The combined flux of the three 
components is $K_{\rm AB,rest-frame}\sim20$, consistent with the $K$-band-redshift 
relation for BCGs (\cite{Whi08}), when extrapolated to $z=2.07$. This suggests that we 
might be witnessing the early stages of the assembly of a brightest cluster galaxy 
through merging (\cite{deL07}).\\

\begin{figure*}
\centering
\includegraphics[width=0.49\textwidth]{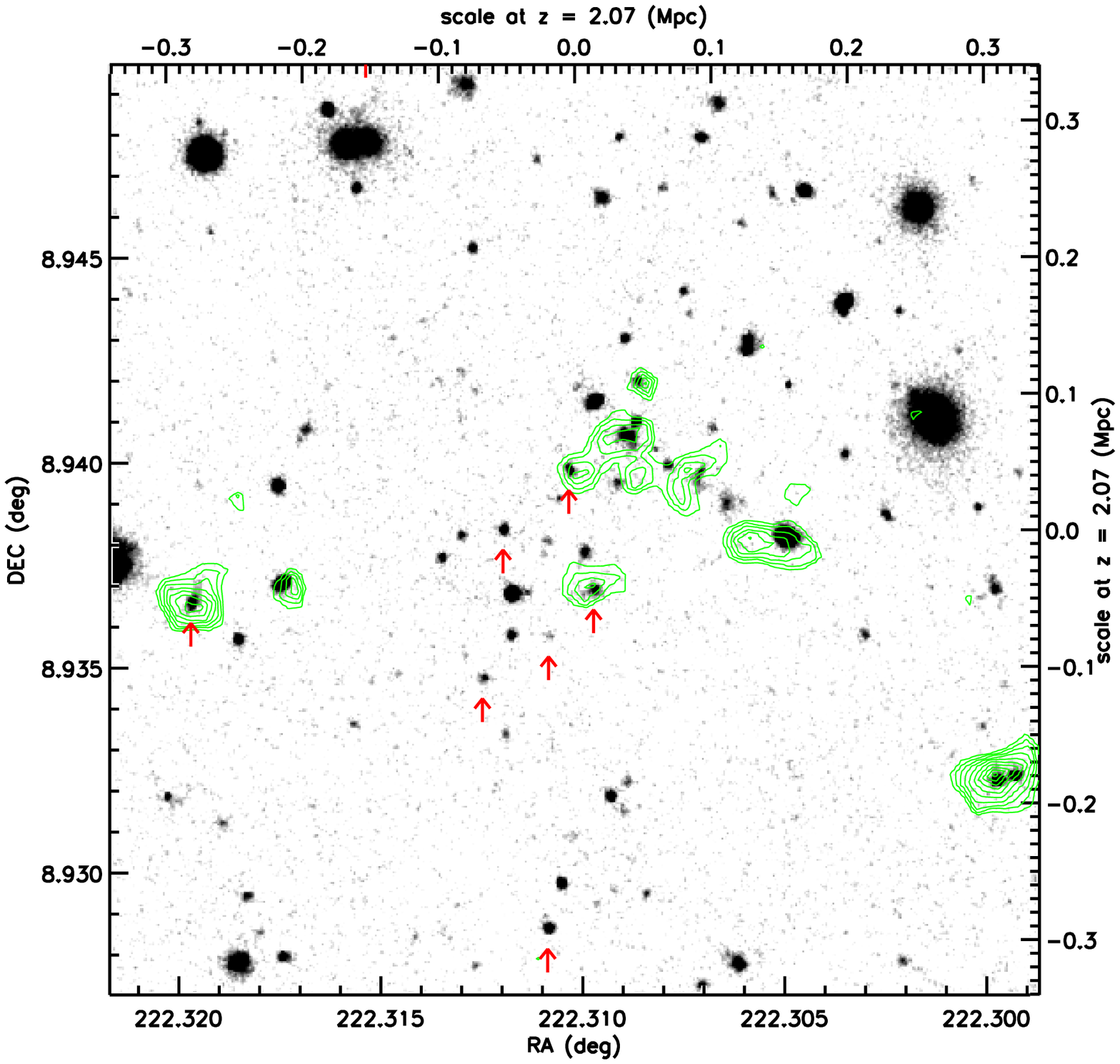}
\caption{$K_s$-band image of the 1.4'$\times$1.4' field centred on the galaxy cluster, as 
in Figs. \ref{fig:rgb1} and \ref{fig:rgb2}, the green contours showing the fluxes of the 
sources detected at 5$\sigma$ or more in the MIPS 24~$\mu$m image. The red arrows show the 
positions of red ($Y-K_s>2$) galaxies with rising IRAC fluxes ($[5.8]-[8.0]>0$), suggesting 
the possible presence of obscured AGN activity.}
\label{fig:mips}
\end{figure*}

We note however that some emission is seen at 24~$\mu$m in and around the cluster centre. 
Several sources are detected at 5$\sigma$ in the core (Fig. \ref{fig:mips}), some of them 
associated with red galaxies, including the AGN and the ``proto-BCG'' triplet.
The PSF-fitted fluxes of these objects are 100--150~$\mu$Jy which, if due to star 
formation, would correspond to ULIRG-like luminosities of L$_{IR}\sim1.5-2\times10^{12}$~L$_{\odot}$ 
and imply very active starbursts with star formation rates of $>150$~M$_{\odot}$ 
yr$^{-1}$, in apparent contradiction with the elliptical-like morphology and compact cores. 
In the case of the ``proto-BCG'' galaxy group, the MIPS PSF is too large (6'') to determine 
which of the three galaxies is a 24~$\mu$m emitter. On the other hand, the 24~$\mu$m emission 
(which at $z\sim2$ corresponds to 8~$\mu$m rest-frame) might be due to extremely obscured 
AGN activity (\cite{Fio09,Dad07b}), an interpretation supported by the presence among the 
MIPS 24~$\mu$m sources of the Chandra-detected AGN. 
Stacking the Chandra data at the position of the five other 24~$\mu$m detected red 
galaxies within the central 1', using a 5'' aperture, we find an excess of photons in the 
hard band at 2.3$\sigma$ significance with respect to the distribution of counts in a thousand 
samples of five random background (i.e. chosen to be at least 5'' away from the nearest $K_s$ 
selected object) positions each. As reported in Section \ref{X}, we find no such excess in the 
soft band. As we might include X-ray emitting but $K$-undetected sources in the background 
positions, the significance level is likely underestimated. However, distinguishing the relative 
contributions of AGN activity and star formation to the mid-IR flux is not within the 
scope of this work and would require far-IR data.\\

\section{\label{X}XMM-Newton and Chandra imaging}

\begin{figure*}
\centering
\includegraphics[width=0.29\textwidth]{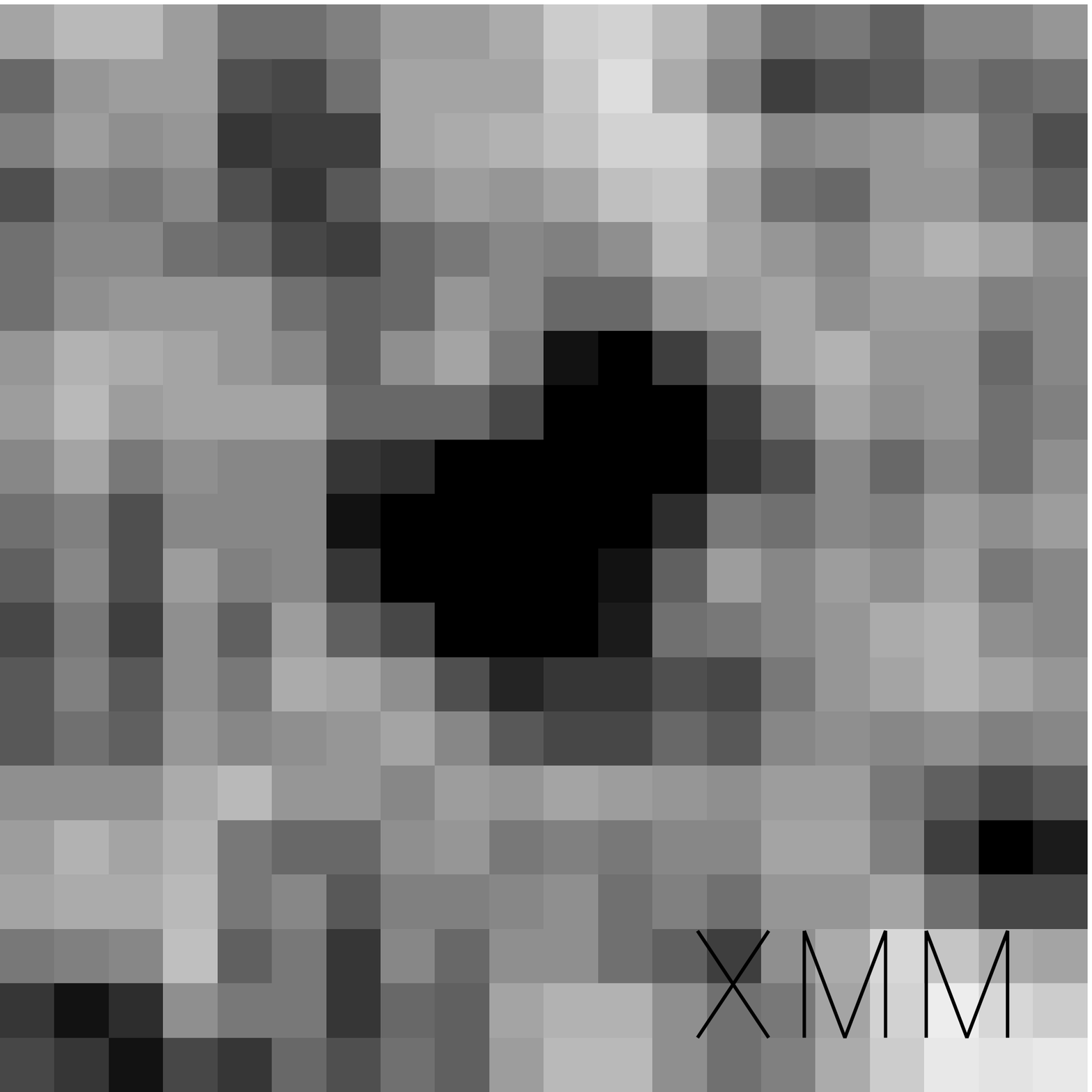}
\includegraphics[width=0.29\textwidth]{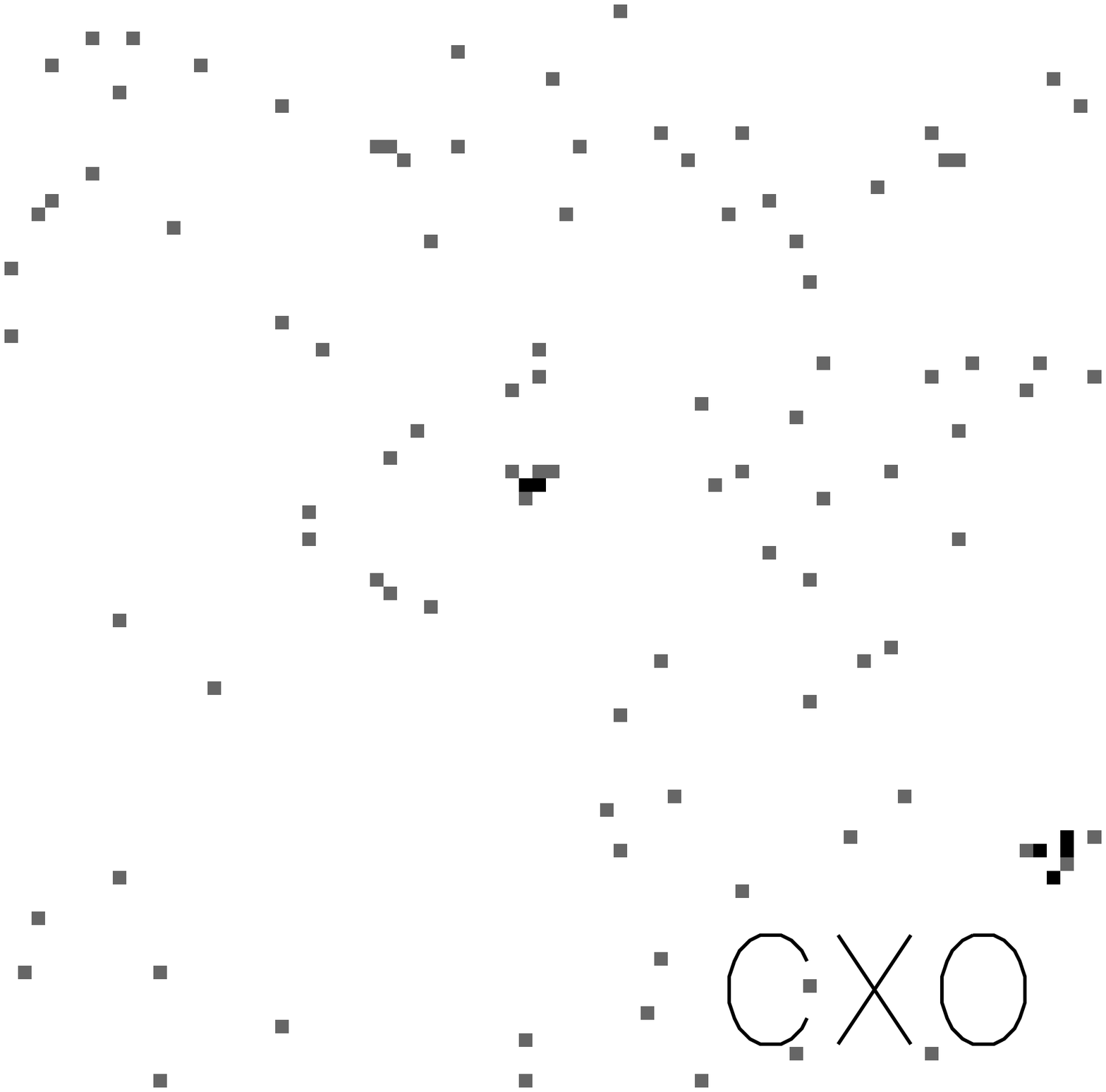}
\includegraphics[width=0.29\textwidth]{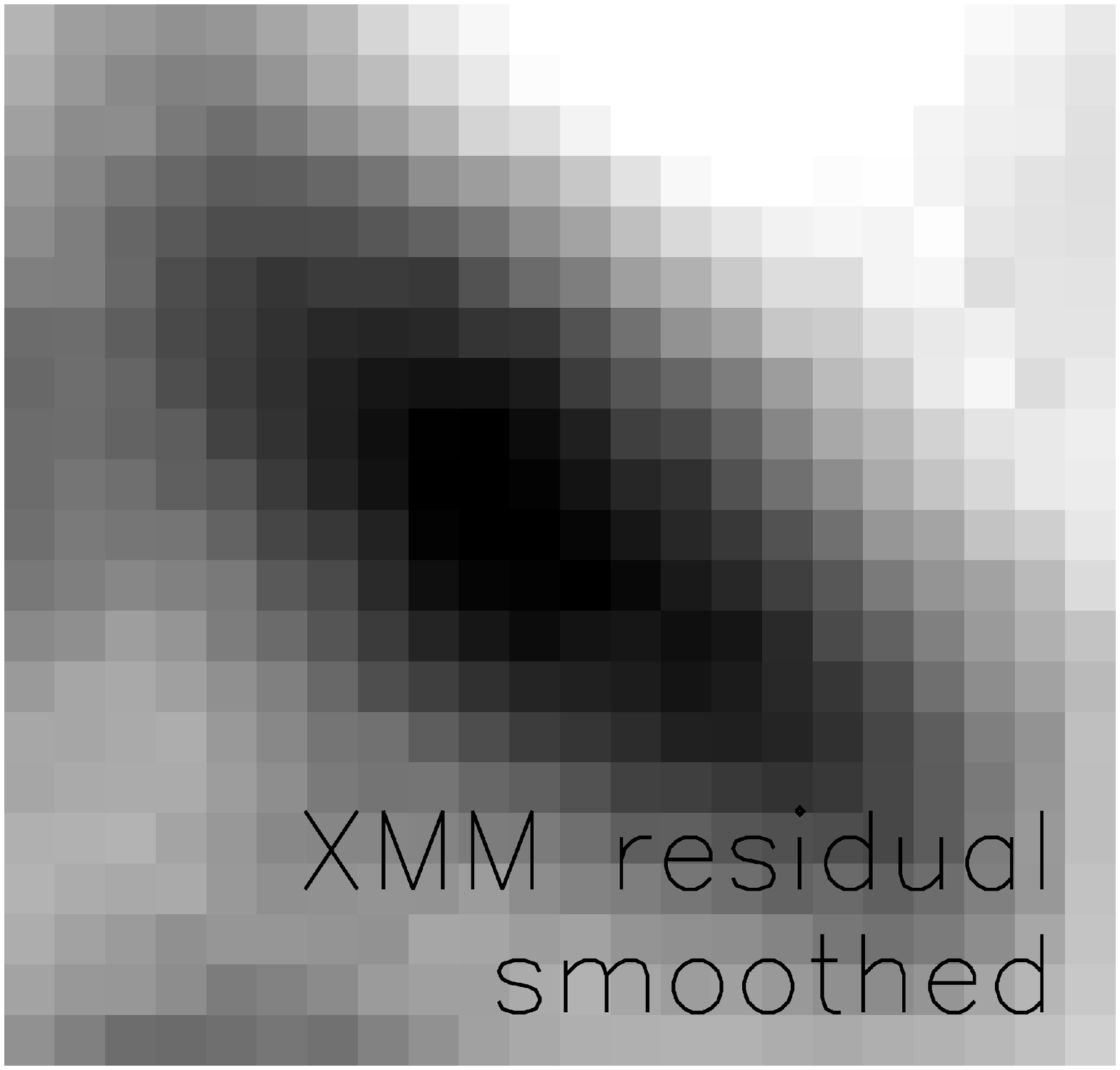}
\caption{Left and middle: soft (0.5-2 keV) X-ray images of the field taken with 
XMM-Newton and Chandra (80ks each), showing, respectively the diffuse emission and 
a point source, an active galactic nucleus corresponding to one of the red galaxies. 
Right: signal-to-noise map of the XMM residual image, after subtraction of the point 
source and smoothing with a FWHM of 16''.}
\label{fig:x}
\end{figure*}

The unambiguous signature of an evolved cluster is the X-ray emission from the ICM, as it 
implies a deep and established potential well. To check for emission from a diffuse 
atmosphere, we looked for extended emission in deep X-ray observations available in the 
field with both the XMM-Newton and Chandra telescopes totalling 80 ks each (\cite{Bru05,Cam09}).
A detection was found in both soft-band (0.5-2 keV) images: the Chandra observation reveals a 
$\sim$1'' point source at the position of one of the red galaxies, while the soft X-ray emission 
seen by XMM is more extended than the 6'' instrument PSF.
To assess the significance of this extended emission, we fitted the XMM emission with a PSF profile 
at the position of the Chandra source and analysed the residual image. After subtraction of the point 
source, we found a residual X-ray emission in the soft band at the $3.5\sigma$ level on scales of 
20--30'', three to five times more extended than the PSF of XMM-Newton. The excess flux over the 
background is $47\pm13$ photons, the error including systematic uncertainties due to the point 
source subtraction. The total flux of this extended emission in the range 0.5--2 keV and over a 
16'' radius is $\sim9.3\times10^{-16}$~erg~s$^{-1}$, consistent with the presence of hot ICM, 
typical of a ``relaxed'' cluster. This is thus faint emission, yet detected at a higher 
significance level than what reported for the $z=1.62$ cluster (\cite{Tan10,Pap10}). As a 
consistency check, we carried out the same analysis on the Chandra data. While we do not 
detect any residual extended emission in the Chandra image, the implied upper limit is nevertheless 
consistent with the XMM result. Fig. \ref{fig:x} shows the XMM and Chandra detections and the 
extended emission after subtraction of the Chandra point source.\\

We emphasise that we kept the X-ray analysis simple and robust. We did not apply 
sophisticated wavelet-based detection techniques (\cite{Fin06}), but only a standard background 
subtraction, point-source subtraction, and photometry over a 16'' radius.  For visualisation 
purposes only, we smoothed the data with a 8'' radius PSF to determine the X-ray contour levels 
shown in Fig. \ref{fig:rgb1}. For an accurate subtraction of point source X-ray emitters, we 
used the deep Chandra observations of the field, exploiting the fact that Chandra has a much 
better spatial resolution than XMM. On the other hand, we note that, as 
the Chandra data were taken one to three years after the XMM observations (Table \ref{tab:phot}), 
AGN variability could have affected our result (\cite{Papa08}). Our measurement should however 
be robust against variability for two strong reasons: first, the fact that the XMM residual 
emission is recovered over an extended area, after point source subtraction, implies that it 
is not an effect of variability. Furthermore, with the position of the AGN known, we fitted 
its flux using the XMM data itself only and found results fully consistent with what is observed 
in Chandra, excluding substantial variability. 
We also considered the possibility that the extended X-ray emission might actually arise 
from several faint AGNs, not individually detected by either XMM or Chandra and spatially 
dispersed in the overdensity. While detailed study of the AGN content in this cluster is 
deferred to a future paper, we carefully investigated which of the galaxies in the structure 
might be hosting an AGN on the basis of either a hard X-ray (2--8 keV) detection, the presence 
of a rising IRAC SED, or of mid-IR excess emission (\cite{Dad07b}). For the latter, we used the 
archival Spitzer MIPS $24\mu$m imaging of the field, which reaches a 5$\sigma$ detection 
limit of about 80~$\mu$Jy. We find that there could be up to 4-5 AGNs in the cluster (as shown in 
Fig. \ref{fig:mips}), in addition to the Chandra point source. However, no soft X-ray photons were 
detected by Chandra at their position. 
Finally, we considered the possibility that the X-ray emission might come from a lower mass 
foreground (e.g. $z\lesssim1$) galaxy group. Such a structure would be more loose and not immediately 
apparent against the backdrop of the main galaxy overdensity but should appear as a distinct peak in 
redshift space. To check for foreground structures, we estimated photometric redshifts for $Y-K_s<2$ 
galaxies without spectroscopic redshifts. For these objects, which include star-forming galaxies as well 
as low redshift ellipticals, we used \cite{CWW} templates, with the addition of a 100~Myr constantly 
star-forming model computed from \cite{M05} templates, and included dust extinction up to 
$E(B-V)=1$. We find no significant secondary structure in the redshift distribution of the 33 galaxies 
(3 spectroscopic and 30 photometric redshifts) within the 1$\sigma$ confidence region of the extended 
X-ray emission. Three blue galaxies with $z_{\rm phot}\sim1.7$ might be associated but their projected 
centre of mass is offset by 15'' with respect to the X-ray centroid.\\

We conclude that, all in all, the only likely explanation is that the extended X-ray emission 
seen by XMM is due to an ICM present in the structure's potential well. 
The presence of an X-ray atmosphere and an evolved galaxy population is consistent with CL J1449+0856
being not a forming proto-cluster but an already mature galaxy cluster comparable to the massive 
structures observed at $z<1.6$.\\

\section{\label{proto}CL J1449+0856 and $\mathbf{z\sim2}$ structures}

While several overdensities of Ly$\alpha$ emitters at $z>2$ have been 
reported (\cite{Pen97,Mil06,St05,Ov06}), CL J1449+0856 is different in 
some key aspects because of the evidence of extended 
X-ray emission from an intra-cluster medium and a centre occupied by 
old passively evolving early-type galaxies. These features make CL 
J1449+0856 much more mature than the high-redshift proto-clusters and 
a unique case among the $z>2$ structures.\\

\subsection{MRC 1138-262}

Among the large galaxy overdensities at $z>2$, that around the massive 
radio galaxy PKS 1138-262 (\cite{Pen97,Mil06}) at $z=2.16$ is the one 
that can be most readily compared to CL J1449+0856. Located at a 
similar redshift, and thus at the same epoch in the history of the 
Universe, it is a massive structure characterised by a giant radio 
galaxy and a host of significantly less massive star-forming 
``satellite'' galaxies (\cite{Mil06}). It has been extensively studied 
photometrically as well as spectroscopically and is understood to be a 
structure still in its formation phase, i.e. a ``proto-cluster''. 
We emphasise that there are significant differences indicating that CL 
J1449+0856 is a very different type of structure, observed at a 
substantially more advanced evolutionary stage than MRC 1138-262.\\

MRC 1138-262 has been observed with Chandra and a number of AGNs 
were detected (\cite{Pen02}) as well as diffuse emission centred on 
the radio galaxy (\cite{Car02}). The latter is however clearly 
associated with the central AGN, which contributes to more than 80\% of 
the total flux, aligned with the radio lobes, suggesting that it is 
due to inverse Compton scattered CMB photons (\cite{Cel04,Fin10}), and 
overall quite different from the typical extended emission from an 
intra-cluster medium (\cite{Car02}). 
On the other hand, the only AGN in the X-ray emission of CL J1449+0856 
does not contribute more than $\sim$50\% of the total observed flux. The 
extended emission itself is not centred on any particular galaxy, but 
clearly spatially associated with the galaxy overdensity, fully 
consistent with a young cluster atmosphere.\\

The proto-cluster MRC 1138-262 was identified as an overdensity of 
Ly$\alpha$ emitters (\cite{Pen00}) and the radio-galaxy itself is 
embedded in a giant emission-line nebula (\cite{Pen97}). The core of 
the proto-cluster is entirely dominated by star-forming galaxies, 
with a few redder galaxies located at the outskirts of the Ly$\alpha$ 
halo (\cite{Ha09}). Of the latter, two have detected H$\alpha$ emission 
(\cite{Do10}), indicating that they are actually actively star-forming 
galaxies. Red galaxies have been found in a broader field surrounding 
the radio-galaxy and Ly$\alpha$ overdensity, a subset of which have 
SEDs and morphologies consistent with a passive population, but the 
claimed overdensity is not concentrated in a putative cluster core 
(\cite{Zi08}), mostly avoiding the central region around the radio-galaxy. 
In contrast, the structure of CL J1449+0856 is very different. Simply 
looking at colour images shows the striking difference between our 
cluster and MRC 1138-262 (compare Fig. 1 in this paper to Fig. 1 in 
\cite{Ha09}, which displays a comparable field of view). Specifically, 
no analogue to the Ly$\alpha$ halo of MRC 1138-262 has been observed 
in CL J1449+0856. None of the red galaxies in the core is seen 
in the rest-frame UV and the space between the galaxies is void of 
light, save for the diffuse X-ray emission. And whereas the core of 
MRC 1138-262 is dominated by star-forming galaxies, the core population 
of CL J1449+0856 is constituted of red, morphologically early-type 
galaxies. This supports our conclusion that, if CL J1449+0856 experienced 
such a star-forming phase as the proto-cluster is undergoing, it 
lies in its past and that the red galaxies are now passively evolving.

Finally, MRC 1138-262, having been discovered as a powerful FRII source 
(\cite{Pen97}), is in practice selected over the whole sky and hence it 
is very difficult to assess its cosmological relevance.

\subsection{JKCS 041}

\cite{And09} claimed the discovery of an evolved, colour-selected galaxy 
cluster at $z\sim1.9$, with extended X-ray emission detected by Chandra. 
From the X-ray luminosity, the authors derive a total mass well above $10^{14}$~M$_{\odot}$. 
The redshift of the structure, JKCS 041, was estimated from photometric redshifts only: 
red-sequence galaxies were observed with FORS2 but proved too faint and no spectroscopic 
information was thus provided to support this identification. The redshift identification of 
this cluster has since been disputed using spectroscopic redshifts from the VIMOS-VLT Deep 
Survey (\cite{LeF05}) and new multi-band photometric redshifts; these data show that JKCS 041 
is not a single high-redshift cluster but a superposition along the line of sight of at least 
two rich galaxy structures, at $z\sim1.1$ and $z\sim1.5$ (\cite{Bie10}). While a $z\sim2$ 
structure could still lurk in the background, this result suggests that the X-ray emission 
originates at $z<1.5$.

\section{\label{disc}Structure and mass of CL J1449+0856}

With an estimated redshift of $z=2.07$ for the cluster, we derive an X-ray luminosity of 
$L_X(0.1-2.4$~keV$)=(7\pm2)\times10^{43}$~erg~s$^{-1}$. On the basis of established $L_X-M$ 
correlations (\cite{Leau10}), the luminosity corresponds to a total mass of 
M$_{200}=(5.3\pm1)\times10^{13}$~M$_{\odot}$, comparable to that of Virgo. The corresponding 
virial radius would be $R_{200}\sim0.37$~Mpc, consistent with the scale of the observed XMM 
emission ($\sim0.4$~Mpc, as shown in Fig. \ref{fig:rgb1}), and a temperature of $kT\sim2$~keV, 
below the detection limit of current SZE facilities (\cite{SZE}). 
We derive a consistent estimate for the total mass from the total stellar mass of the 
red member galaxies in the central 20'', $M^{\star}_{\rm tot}=4.9\times10^{11}$~M$_{\odot}$. 
Using the locally calibrated stellar mass-to-halo mass conversion 
of \cite{Mos10}, we find a halo mass of $M_{\rm halo}=8.0\times10^{13}$~M$_{\odot}$ and a 
corresponding virial radius of 0.42~Mpc (assuming $M_{\rm halo}=$M$_{200}$). We note that 
both estimates of the virial radius are comparable to the size of the overdensity from which 
we selected the red galaxies, 20'' corresponding to $\sim0.17$~Mpc proper at $z=2.07$. 
Furthermore, assuming that the core is in the first order spherical, at $z=2.07$ a sphere 
of radius 0.17~Mpc with a mass of $8\times10^{13}$~M$_{\odot}$ has a mean density of 
$\sim3000\rho_c$ (where the critical density at $z=2.07$ is $\rho_c=8.63\times10^{-29}$~g~cm$^{-3}$), 
significantly above the usual threshold of $\sim178\rho_c$ for virialisation. 
Although this estimate has substantial uncertainties, this might suggest that the core of 
CL J1449+0856 has already collapsed and (at least partially) virialised.
Both total mass estimates might be regarded as lower limits, as the 
$L_X-M$ relation assumes that the intra-cluster gas has fully reached its virial 
temperature, a state that CL J1449+0856 may still be approaching, and our census of 
massive galaxies is incomplete as we considered only the red members in a small 
central region, without including star-forming members or galaxies outside the 
immediate core. The agreement of the two estimates, however, suggests that the faint 
X-ray emission is reliable. It would seem to imply that not only the $M^{\star}-M$ relation 
holds to $z\sim2$ and but that this cluster is, to the first order, already on the $L_X-M$ 
correlation.\\

With these halo mass estimates, we can now use the results of numerical simulations 
from the literature to attempt a prediction of the future growth of CL J1449+0856 
and its final mass at $z=0$. Based on the mass assembly histories of halos in 
the Millennium and Millennium-II simulations (\cite{Fak10}), we estimate that 
CL J1449+0856 would reach a mass of $0.9-1.5\times10^{14}$~M$_{\odot}$ at $z=1.5$ 
and $4.9-8.2\times10^{14}$ at $z=0$. These values are well within the range of 
cluster masses at these respective redshifts. At $z=1-1.5$, CL J1449+0856 would 
be about one fourth as massive as the most massive $z>1$ clusters (\cite{Bro10,Ro09}) 
and reach a $z=0$ mass comparable to that of the Coma cluster (\cite{Ku07}).\\

Assuming Gaussian initial conditions and concordance cosmology with 
$\sigma_8=0.8$ (\cite{SZE}), the probability of finding a dark matter halo 
with $z\ge 2.07$ and a mass greater than $5\times10^{13}$~M$_{\odot}$ in 
the survey area of $400$~arcmin$^2$ is $3.3\times10^{-2}$. Allowing 
some time before the dark matter halo formation and observation (\cite{Ji09}), 
and the gas to settle in the deep potential well, would 
further lower this probability, e.g. down to $1.0\times10^{-2}$ for 
$z_f=2.30$ (0.3~Gyr before the observations). Lowering $\sigma_8$ to 
0.77 would further reduce these numbers by roughly a factor of 3. 
Although based on a single object detection, our finding suggests an 
excess of massive high-redshift structures, consistent with the 
constraints derived from the existence of a very massive cluster at 
$z=1.39$ (\cite{Jee09,Ji09}). However, we note that no obvious similarly 
high-redshift candidate structure is detected in the 2 deg$^2$ COSMOS 
field (\cite{Tan11,Sa11}).

\section{\label{sum}Summary}

We have discovered a remarkable structure whose properties are consistent with 
it being a mature cluster at $z=2.07$. This structure was selected as an 
overdensity of sources with IRAC colours satisfying $[3.6]-[4.5]>0$. Deep 
follow-up observations with Subaru and the VLT revealed a strong overdensity 
of galaxies with $Y-K_s$ colours consistent with a passive population at 
$z\gtrsim2$. We also obtained high resolution HST/NICMOS and Keck AO images, 
which revealed that the red galaxies have elliptical-like morphologies and 
compact cores.\\
From the VLT VIMOS and FORS2 spectra of $sBzK$ galaxies around the core, we 
estimated a redshift of $z=2.07$ for the structure. We estimated photometric 
redshifts from the 12-band SEDs of the red galaxies, whose distribution peaks 
at $z=2.05$, in agreement with the spectroscopic redshifts.\\
Using XMM-Newton and Chandra observations of the field, we found an extended 
soft X-ray emission at the 3.5$\sigma$ confidence level, at the position of 
the galaxy overdensity. The observed X-ray luminosity and the galaxy mass 
content of the core imply a total halo mass of 5--8$\times10^{13}$~M$_{\odot}$.\\

Our results show that virialised clusters with detectable X-ray emission 
and a fully established early-type galaxy content were already in 
place at $z>2$, when the Universe was only $\sim3$~Gyr old. While it took us 
several years of observations to confirm this structure, upcoming facilities 
like JWST and future X-ray observatories should be able of routinely find and 
study similar clusters, unveiling their thermodynamic and kinematic structure 
in detail. The census of $z>2$ structures similar to CL J1449+0856 will subject 
the assumed Gaussianity of the primordial density field to a critical check.

\begin{acknowledgements}
This work is based on data collected at the Subaru Telescope, which is operated by the National 
Astronomical Observatory of Japan; on observations made with ESO telescopes at the Paranal 
Observatory, under programmes 072.A-0506 and 381.A-0567; and on observations made with the 
NASA/ESA Hubble Space Telescope, which is operated by the Association of Universities for Research 
in Astronomy, Inc., under NASA contract NAS 5-26555. Support for programme \#11174 was provided 
by NASA through a grant from the Space Telescope Science Institute, which is operated by the 
Association of Universities for Research in Astronomy, Inc., under NASA contract NAS 5-26555. 
Some of the data presented herein were obtained at the W.M. Keck Observatory, which is operated 
as a scientific partnership among the California Institute of Technology, the University of 
California and the National Aeronautics and Space Administration.
We acknowledge funding ERC-StG-UPGAL-240039, ANR-07-BLAN-0228 and ANR-08-JCJC-0008 and Alvio 
Renzini acknowledges financial support from contract ASI/COFIS I/016/07/0. This work is partially 
supported by a Grant-in-Aid for Science Research (No. 19540245) by the Japanese Ministry of 
Education, Culture, Sports, Science and Technology. We thank Romain Teyssier for his help with 
the cosmological calculations, Monique Arnaud, David Elbaz and Piero Rosati for discussions.
\end{acknowledgements}


\begin{thebibliography}{1}

\bibitem[Andreon et al. 2009]{And09}Andreon, S., Maughan, B., Trinchieri, 
G., Kurk, J., 2009, A\&A 507, 147-157
\bibitem[Beers, Flynn \& Gebhardt 1990]{Be90}Beers, T.C., Flynn, K., Gebhardt, 
K., 1990, AJ 100, 32-46
\bibitem[Bertin \& Arnouts 1996]{BA96}Bertin, E., Arnouts, S., 1996, A\&AS 117, 
393-404
\bibitem[Bielby et al. 2010]{Bie10}Bielby, R.M. et al., 2010, 
\emph{http://arxiv.org/abs/1007.5236}
\bibitem[Binggeli et al. 1987]{Bing87}Binggeli, B., Tammann, G.A., Sandage, 
A., 1987, AJ 94, 251-277
\bibitem[Blakeslee et al. 2003]{Bla03}Blakeslee, J.P. et al., 2003, ApJ 596, 
L143-L146
\bibitem[Boselli \& Gavazzi 2006]{Bo06}Boselli, A., Gavazzi, G., 2006, PASP 118, 
517-559
\bibitem[Brodwin et al. 2010]{Bro10}Brodwin, M. et al., 2010, ApJ 721, 90-97
\bibitem[Brusa et al. 2005]{Bru05}Brusa, M. et al., 2005, A\&A 432, 69-81
\bibitem[Campisi et al. 2009]{Cam09}Campisi, M.A. et al., 2009, A\&A 501, 485-494
\bibitem[Carilli et al. 2002]{Car02}Carilli, C.L. et al., 2002, ApJ 567, 781-789
\bibitem[Cay\'on et al. 2010]{Ca10}Cay\'on, L., Gordon, C., Silk, J., 2010, 
\emph{http://arxiv.org/abs/1006.1950}
\bibitem[Celotti \& Fabian 2004]{Cel04}Celotti, A., Fabian, A.C., 2004, MNRAS 353, 
523-528
\bibitem[Chabrier 2003]{Cha03}Chabrier, G., 2003, PASP 115, 763-795
\bibitem[Coleman, Wu \& Weedman 1980]{CWW}Coleman, G.D., Wu, C.-C., Weedman, 
D.W., 1980, ApJS 43, 393-416
\bibitem[Coles \& Lucchin 1995]{CL95}Coles, P., Lucchin, F. 1995, Cosmology: 
The Origin and Evolution of Cosmic Structure, Wiley
\bibitem[Daddi et al. 2000]{Dad00}Daddi, E. et al., 2000, A\&A 361, 535-549
\bibitem[Daddi et al. 2004]{Dad04}Daddi, E. et al., 2004, ApJ 617, 746-764
\bibitem[Daddi et al. 2005]{Dad05}Daddi, E. et al., 2005, ApJ 626, 680-697
\bibitem[Daddi et al. 2007a]{Dad07a}Daddi, E. et al., 2007, ApJ 670, 156-172
\bibitem[Daddi et al. 2007b]{Dad07b}Daddi, E. et al., 2007, ApJ 670, 173-189
\bibitem[Daddi et al. 2009]{Dad09}Daddi, E. et al., 2009, ApJ 694, 1517-1538
\bibitem[De Lucia \& Blaizot 2007]{deL07}De Lucia, G., Blaizot, J., 2007, 
MNRAS 375, 2-14
\bibitem[Demarco et al. 2010]{De10}Demarco, R. et al., 2010, ApJ 725, 1252-1276
\bibitem[Doherty et al. 2010]{Do10}Doherty, M. et al., 2010, A\&A 509, A83
\bibitem[Elbaz et al. 2007]{El07}Elbaz, D. et al., 2007, A\&A 468, 33-48
\bibitem[Fakhouri et al. 2010]{Fak10}Fakhouri, O., Ma, C.-P., Bolyan-Kolchin, 
M., 2010, MNRAS 406, 2267-2278
\bibitem[Finoguenov et al. 2006]{Fin06}Finoguenov, A. et al., 2007, ApJS 172, 
182-195
\bibitem[Finoguenov et al. 2010]{Fin10}Finoguenov, A. et al., 2010, MNRAS 403, 
2063-2076
\bibitem[Fiore et al. 2007]{Fio08}Fiore, F. et al., 2008, ApJ 672, 94-101
\bibitem[Fiore et al. 2009]{Fio09}Fiore, F. et al., 2009, ApJ 693, 447-462
\bibitem[Francis et al. 1996]{Fra96}Francis, P.J. et al., 1996, ApJ 457, 490-499
\bibitem[Gladders \& Yee 2000]{Gla00}Gladders, M.D., Yee, H.K.C., 2001, AJ 120, 
2148-2162
\bibitem[Hatch et al. 2009]{Ha09}Hatch, N.A. et al., 2009, MNRAS 395, 114-125
\bibitem[Haiman et al. 2001]{Hai01}Haiman, Z., Mohr, J.J., Holder, G.P., 2001, 
ApJ 553, 545-561
\bibitem[Hashimoto et al. 1998]{Has98}Hashimoto, Y., Oemler, A., Jr., Lin, H., 
Tucker, D.L., 1998, ApJ 499, 589-599
\bibitem[Hayashi et al. 2010]{Hay10}Hayashi, M. et al., 2010, MNRAS 402, 1980-1990
\bibitem[Henry et al. 2010]{He10}Henry, J.P. et al., 2010, 
\emph{http://arxiv.org/abs/1010.0688}
\bibitem[Hilton et al. 2010]{Hi10}Hilton, M. et al., 2010, ApJ 718, 133-147
\bibitem[Hwang \& Park 2009]{Hwa09}Hwang, H.S., Park, C., 2009, ApJ 700, 
791-798
\bibitem[Ilbert et al. 2006]{Il06}Ilbert, O. et al., 2006, A\&A 457, 841-856
\bibitem[Jee et al. 2009]{Jee09}Jee, M.J. et al., 2009, ApJ 704, 672-686
\bibitem[Jimenez \& Verde 2009]{Ji09}Jimenez, R., Verde, L., 2009, PhRvD 80, 
127302
\bibitem[Kinney et al. 1996]{Kin96}Kinney, A.L. et al., 1996, ApJ 467, 38-60
\bibitem[Kodama \& Arimoto 1997]{KA97}Kodama, T., Arimoto, N., 1997, A\&A 320, 
41-53
\bibitem[Kodama et al. 2007]{Ko07}Kodama, T. et al., 2007, MNRAS 377, 1717-1725
\bibitem[Koekemoer et al. 2002]{Koe02}Koekemoer, A.M., Fruchter, A.S., Hook, 
R.N., Hack, W., 2002, The 2002 HST Calibration Workshop: Hubble after the 
installation of the ACS and the NICMOS Cooling System, 337-340
\bibitem[Kong et al. 2006]{Kong06}Kong, X. et al., 2006, ApJ 638, 72-87
\bibitem[Kubo et al. 2007]{Ku07}Kubo, J.M. et al., 2007, ApJ 671, 1466-1470
\bibitem[Kurk et al. 2009]{Kur09}Kurk, J. et al. 2009, A\&A 504, 331-346
\bibitem[Leauthaud et al. 2010]{Leau10}Leauthaud, A. et al., 2010, ApJ 709, 
97-114
\bibitem[Le F\`evre et al. 2005]{LeF05}Le F\`evre, O. et al., 2005, A\&A 439, 
845-862
\bibitem[Magee, Bouwens \& Illingworth 2007]{Mag07}Magee, D.K., Bouwens, R.J., 
Illingworth, G.D., 2007, ASPC 376, 261-264
\bibitem[Maraston 2005]{M05}Maraston, C., 2005, MNRAS 362, 799-825
\bibitem[Maraston et al. 2006]{M06}Maraston, C. et al., 2006, ApJ 652, 85-96
\bibitem[Miley et al. 2006]{Mil06}Miley, G.K. et al., 2006, ApJ 650, L29-L32
\bibitem[Moster et al. 2010]{Mos10}Moster, B.P. et al., 2010, ApJ 710, 903-923
\bibitem[Mullis et al. 2005]{Mul05}Mullis, C.R. et al., 2005, ApJ 623, L85-L88
\bibitem[Overzier et al. 2006]{Ov06}Overzier, R. et al., 2006, ApJ 637, 58-73
\bibitem[Overzier et al. 2008]{Ov08}Overzier, R. et al., 2008, ApJ 673, 143-162
\bibitem[Papadakis et al. 2008]{Papa08}Papadakis, I.E., Chatzopoulos, E., 
Athanasiadis, D., Markowitz, A., Georgantopoulos, I., 2008, A\&A 487, 475-483
\bibitem[Papovich et al. 2010]{Pap10}Papovich, C. et al., 2010, ApJ 716, 
1503-1513
\bibitem[Park \& Hwang 2009]{Par09}Park, C., Hwang, H.S., 2009, ApJ 699, 1595-1609
\bibitem[Patel et al. 2009]{Pa09}Patel, S.G., Holden, B.P., Kelson, D.D., 
Illingworth, G.D., Franx, M., 2009, ApJ 705, L67-L70
\bibitem[Peacock 1999]{Pea99}Peacock, J.A., 1999, Cosmological Physics, Cambridge 
Univ. Press
\bibitem[Peebles 1993]{Pee93}Peebles, P.J.E., 1993, Physical Cosmology, Princeton 
Univ. Press
\bibitem[Peng et al. 2002]{Peng02}Peng, C.Y., Ho, L.C., Impey, C.D., Rix, H.-W., 
2002, AJ 124, 266-293
\bibitem[Peng et al. 2010]{Pe10}Peng, Y.-J. et al., 2010, ApJ 721, 193-221
\bibitem[Pentericci et al. 1997]{Pen97}Pentericci, L., Roettgering, H.J.A., 
Miley, G.K., Carilli, C.L., McCarthy, P., 1997, A\&A 326, 580-596
\bibitem[Pentericci et al. 2000]{Pen00}Pentericci, L. et al., 2000, A\&A 361, 
L25-L28
\bibitem[Pentericci et al. 2002]{Pen02}Pentericci, L. et al., 2002, A\&A 396, 
109-115
\bibitem[Pierre et al. 2003]{Pie03}Pierre, M. et al., 2003, JCAP 9, 11
\bibitem[Ponman et al. 1999]{Pon99}Ponman, T.J., Cannon, D.B., Navarro, J.F., 
1999, Nature 397, 135-137
\bibitem[Postman et al. 2005]{Pos05}Postman, M. et al., 2005, ApJ 623, 721-741
\bibitem[Press \& Schechter 1974]{PS74}Press, W.H., Schechter, P., 1974, ApJ 
187, 425-438
\bibitem[Rettura et al. 2010]{Re10}Rettura, A. et al., 2010, ApJ 709, 512-524
\bibitem[Romer et al. 2001]{Rom01}Romer, A.K., Viana, P.T.P., Liddle, A.R., Mann, 
R.G, 2001, ApJ 547, 594-608
\bibitem[Rosati et al. 1998]{Ro98}Rosati, P., della Ceca, R., Colin, N., Giacconi, 
R., 1998, ApJ 492, L21-L24
\bibitem[Rosati et al. 2002]{Ro02}Rosati, P., Borgani, S., Norman, C., 2002, 
ARA\&A 40, 539-577
\bibitem[Rosati et al. 2009]{Ro09}Rosati, P. et al., 2009, A\&A 508, 583-591
\bibitem[Salvato et al., in prep.]{Sa11}Salvato, M. et al., 2011, in preparation
\bibitem[S\'ersic 1963]{Ser63}S\'ersic, J.-L., Boletin de la Asociacion Argentina 
de Astronomia 6, 41
\bibitem[Schuecker et al. 2002]{Schu02}Schuecker, P., B\"ohringer, H., Collins, 
C.A., Guzzo, L., 2002, A\&A 398, 867-877
\bibitem[Schuecker et al. 2003]{Schu03}Schuecker, P., B\"oringer, H., Collins, 
C.A., Guzzo, L., 2003, A\&A 398, 867-877
\bibitem[Shapley et al. 2003]{Shap03}Shapley, A.E., Steidel, C.C., Pettini, M., 
Adelberger, K.L., 2003, ApJ 588, 65-89
\bibitem[Scodeggio et al. 2005]{Sco05}Scodeggio, M. et al., 2005, PASP 117, 1284-1295
\bibitem[Stanford et al. 2006]{Sta06}Stanford, S.A. et al., 2006, ApJ 646, 
L13-L16
\bibitem[Steidel et al. 2000]{St00}Steidel, C.C. et al., 2000, ApJ 532, 170-182
\bibitem[Steidel et al. 2005]{St05}Steidel, C.C. et al., 2005, ApJ 626, 44-50
\bibitem[Tanaka et al. 2010]{Tan10}Tanaka, M., Finoguenov, A., Ueda, Y. A, 2010, 
ApJ 716, L152-L156
\bibitem[Tanaka et al., in prep.]{Tan11}Tanaka, M. et al., 2011, in preparation
\bibitem[Toft et al. 2007]{Tof07}Toft, S. et al., 2007, ApJ 671, 285-302
\bibitem[Tran et al. 2010]{Tra10}Tran, K.-V.H. et al., 2010, ApJ 719, L126-L129
\bibitem[Vanderlinde et al. 2010]{SZE}Vanderlinde, K. et al., 2010, ApJ 722, 
1180-1196
\bibitem[van der Wel et al. 2010]{vdW10}van der Wel, A., Bell, E.F., Holden, 
B.P., Skibba, R.A., Rix, H.-W., 2010, ApJ 714, 1779-1788
\bibitem[Voit 2005]{Voit05}Voit, G.M., 2005, RvMP 77, 207-258
\bibitem[Whiley et al. 2008]{Whi08}Whiley, I.M. et al., 2008, MNRAS 387, 
1253-1263
\bibitem[Wilson et al. 2008]{Wi08}Wilson, G. et al., 2008, ASPC 381, 210-215
\bibitem[Zirm et al. 2008]{Zi08}Zirm, A.W. et al., 2008, ApJ 680, 224-231

\end{thebibliography}
\end{document}